\documentstyle[12pt]{article}
\renewcommand{\theequation}{\thesection.\arabic{equation}}

\def\e{{\rm e}}
\def\f{\tilde{f}}

\def\D{{\cal D}}

\def\I{{\cal I}}
\def\J{{\cal J}}
\def\M{{\cal M}}
\def\O{{\rm O}\!}

\def\S{{\cal S}}
\def\Z{{\cal Z}}
\def\d{{\rm d}}
\def\to{\rightarrow}
\def\as{\alpha_s}
\def\qes{Q_{\rm ES}^2}
\def\shat{\hat{s}}
\def\dc{d_{\rm cut}}
\def\Sp{{\rm Split}^{\rm tree}}
\def\smin{s_{\rm min}}

\def\soft{{\rm soft}}
\def\coll{{\rm coll}}
\def\fin{{\rm fin}}
\def\ini{{\rm ini}}
\def\alphamax{\alpha_{\rm max}}
\def\deltamax{\delta_{\rm max}}

\def\cG{c_\Gamma}

\newcommand{\la}{\langle}
\newcommand{\ra}{\rangle}
\def\vspaceinarray{\nonumber ~&~&~\\}
\newcommand{\Nc}{N_c}
\newcommand{\Nf}{N_f}

\def\AP{Altarelli-Parisi }
\def\NLO{next-to-leading order }
\def\A#1#2{\la#1#2\ra}
\def\B#1#2{[#1#2]}
\def\s#1#2{s_{#1#2}}

\def\P{\tilde{P}}

\def\del#1{\lower.25em\hbox{\LARGE $\times$}\kern -1em #1 }
\def\st#1{\lower.75em\hbox{$|_{#1}$} }

\newcommand{\ve}{\varepsilon}
\newcommand{\beq}{\begin{equation}}
\newcommand{\eeq}{\end{equation}}
\newcommand{\beqn}{\begin{eqnarray}}
\newcommand{\eeqn}{\end{eqnarray}}
\newcommand{\beqns}{\begin{eqnarray*}}
\newcommand{\eeqns}{\end{eqnarray*}}

\def\Am{{\cal A}}
\def\nn{\nonumber}

\def\ms{$\overline{{\rm MS}}$}

\begin{document}

\begin{titlepage}
\vspace*{-2cm}
\begin{flushright}
KLTE-DTP/96-1\\
February, 1996 \\
\end{flushright}
\vskip .5in
\begin{center}
{\large\bf
Calculation of QCD jet cross sections at \NLO}\\
\vspace*{1.5cm}
{\large Zolt\'an Nagy and 
Zolt\'an Tr\'ocs\'anyi}\footnote{Zolt\'an Magyary fellow,
e-mail: zoltan@dtp.atomki.hu} \\
\vskip 0.2cm
Department of Theoretical Physics, KLTE, \\
Debrecen, Hungary \\
\vskip 2cm
%draft
\vskip 1cm
\end{center}

\begin{abstract}
\noindent A general method for calculating \NLO cross sections in
perturbative QCD is presented.  The algorithm is worked out for
calculating $N$-jet cross sections in hadron-hadron collisions.
The generalization of the scheme to performing caclulations for
other QCD process, such as
electron-positron annihilation or in deep inelastic scattering is also
straightforward. As an illustration several three-jet cross section
distributions in electron-positron annihilation, calculated using the
algorithm, are presented.
\bigskip

\end{abstract}
\end{titlepage}
\setcounter{footnote}{0}

\section{Introduction}
\setcounter{equation}{0}

It is well known that owing to large scale dependence any result of a
leading order calculation in perturbative QCD can be regarded only as 
a one parameter fit to the data, but not a real theoretical prediction.
The scale dependence is expected to decrease substantially with the
inclusion of \NLO corrections, which in a sense fixes the scale.
Consequently, the calculation of cross sections at \NLO in perturbative 
QCD is highly desirable. During the last sixteen years a good number 
of such calculations have been performed. A common feature of these
calculations is that they use the specific, usually simple kinematics
of the problem in order to achieve the analytic cancelation of infrared
singularities.  This specialization brings an element 
of art into \NLO calculations, which for more complex problems like the 
calculation of four-jet production in $\e^+\e^-$ annihilation or 
three-jet production in hadron-hadron collisions becomes an obstacle
to overcome. One would rather like to apply a sort of ``standard
technique'' in order to discriminate the problems of book-keeping from
those of the theory. The first steps into the direction of drawing a
standard picture were done by Ellis, Kunszt and Soper in refs.\
\cite{EKSjets,KSjets} and by Giele, Glover and Kosower in refs.\
\cite{GGepem,GGKpp}. The common characteristics of these works is the
recognition that the factorization properties of the QCD amplitudes or
squared matrix elements \cite{factorization} as well as that of the
phase space in the limits when one particle becomes soft, or an external
pair becomes collinear can be utilized for devising a universal scheme for
the calculation of any infrared safe physical cross section at \NLO in
perturbative QCD.\footnote{More recently an elegant scheme has been
outlined by Catani and Seymour in ref.\ \cite{CSdipol}.}

The algorithm developed in refs.\ \cite{EKSjets,KSjets} --- called the 
subtraction method --- was applied for calculating inclusive one-jet 
\cite{EKSincl} as well as for two-jet \cite{EKS2jet} production at 
next-to-leading order. The same algorithm however, cannot be directly
applied to more complex cases --- like the ones mentioned above~---,
because the evaluation of certain integrals used the specific
$2\to 2$ kinematics of the problem considered and also because the
algorithm relies on single singular decomposition of the squared matrix
element. The latter is not a problem in principle, however, the last
decade has proved that one has to use helicity amplitudes both at
tree (see eg.\ \cite{MP90}) as well as at loop level
\cite{BDK5g,KST4q1g,BDK2q3g} in order to obtain higher order results
and it is rather cumbersome to square these amplitudes analytically
and perform the single singular factoring of the squared matrix elements.
The subtraction method has been generalized to the calculation
of three-jet cross sections in hadron collisions in a recent paper by
Frixione, Kunszt and Signer \cite{FKSjets}. In this paper however, the
physical quantity --- the ``measurement function'' --- was used for
coping with the problem of single singular factoring, which makes the
generalization to different types of cross sections than the one
discussed in the paper non-trivial.

The algorithm of refs.\ \cite{GGepem,GGKpp} --- called the slicing
method --- avoid the above obstacles offering a general scheme, but
at the price of introducing an unphysical parameter $\smin$ and 
calculating the result to O($\smin$) accuracy. In principle $\smin$ can 
be chosen infinitesimal, thus an exact result can be recovered. However,
in practice the choice of a very small $\smin$ adversely affects the
numerical convergence of the Monte Carlo integrals and one has to carry
out a balancing procedure between the error of the Monte Carlo
integration and the one introduced by the choice of finite $\smin$ in 
order to minimize the theoretical error. This balancing procedure
can be inconvenient in those cases when the matrix elements are
complicated and their numerical evaluation is time consuming.

The aim of the present paper is to provide a simple generalization of the 
subtraction method that can be used for calculating any infrared safe 
physical quantity at \NLO in perturbative QCD if the required tree and
one-loop level helicity amplitudes are known. In order to minimize the 
theoretical error of the Monte Carlo integrals, we apply important
sampling. There are two ways of achieving efficient important sampling.
One is when the integrand is decomposed into single singular factors and
the important sampling is performed in the variable controlling the
singularity. As stated above, however, single singular factoring is
better avoided. The second possibility, that we apply in this paper, is
a decomposition of the phase space into regions, where the integrand can
become singular due to the vanishing of only one Lorentz invariant of
the external momenta. This decomposition can be done quite generally,
without any reference to the squared matrix element, or to the physical
quantity being calculated. We describe the algorithm in detail for the
case of hadron collisions, which is the most general case one can
encounter. Algorithms for other processes can be obtained by leaving out
certain terms as it will be explained later.

In section 2, we discuss how infrared safe cross sections can be
calculated in perturbative QCD. In the
following sections we describe the cancelation scheme in detail.
The scheme is based upon the soft and collinear factorization
properties of the squared matrix elements of QCD and that of the phase
space. The singularity structure of the one-loop amplitudes for QCD
processes involving arbitrary number of external partons has been discussed
in ref.\ \cite{KSTsing}. Using those results it is not difficult to find
the universal structure of the singularities in the \NLO matrix
element of the virtual corrections. That universal structure has already
been given in refs.\ \cite{KSjets,KST2to2}. For the sake of completeness as
well as for setting some of the notation, we recall the necessary 
formulas in section 3. 

Section 4 contains the essence of our algorithm. Here we discuss 
the decomposition of the
phase space, the singularity structure of the real corrections. We
describe how the phase space is generated to achieve the necessary
important sampling. We define local soft and collinear
subtraction terms that make the integral of the real corrections over the
$N+1$ particle phase space finite. The explicit expression for this
finite integral is also given. We integrate out the variables of the soft
or collinear particle analytically in sections 5 and 6. We show that the
remaining expression has the form of the $2\to N$ integrals (like the 
Born and virtual corrections), so they can be combined and the analytic
cancelation of the infrared divergences is demonstrated. The remaining
finite $2\to N$ integral is explicitly given. Section 7 contains some
sample results for three-jet cross section calculation in $e^+e^-$
annihilation. We conclude in section 8. The appendix is a collection of
the analytic integrals that were used in the main text for the
demonstration of the cancelation of the infrared divergences.

%\newpage

\section{Infrared safe cross sections at \NLO }
\setcounter{equation}{0}

At order $\as^{(N+1)}$, one calculates cross sections, with infrared
divergences controlled using dimensional regularization in $d=4-2\ve$
dimensions, for the two incoming hadrons to collide and produce either
$N$ or $N+1$ final state partons. According to the factorization
theorem, the \NLO infrared safe physical cross section is a sum of two
integrals,
\beq
\label{physical}
\sigma = \I[2\to N] + \I[2\to N+1],
\eeq
where 
\beqn
\label{2toN}
&&\I[2\to N] =
\sum_{a_A,a_B,a_1,\ldots,a_N}
\int \d x_A \f_A (a_A,x_A)\int \d x_B \f_B (a_B,x_B)
\\ \nn &&\qquad\qquad\times  
\frac{1}{2 \hat{s}} \int \d \Gamma^{(N)}(p_1^\mu,\ldots,p_N^\mu)
\la |\M (2\to N)|^2 \ra \S_N(p_1^\mu,\ldots,p_N^\mu)
\eeqn
and 
\beqn
\label{2toN+1}
&&\I[2\to N+1] =
\sum_{a_A,a_B,a_1,\ldots,a_{N+1}}
\int \d x_A f_A(a_A,x_A)\int \d x_B f_B(a_B,x_B)
\\ \nn &&\qquad\qquad\times
\frac{1}{2 \hat{s}} \int \d \Gamma^{(N+1)}(p_1^\mu,\ldots,p_{N+1}^\mu)
\la |\M (2\to {N+1})|^2 \ra \S_{N+1}(p_1^\mu,\ldots,p_{N+1}^\mu).
\eeqn
In these equations $\hat{s} = x_A x_B s$. In the phase space measures,
\beq
\d \Gamma^{(n)}(p_1^\mu,\ldots,p_n^\mu) =
\frac{1}{n!} \prod_{i=1}^n 
\left(\frac{\mu^{2\ve}\d^{d-1}{\bf p}_i}{(2\pi)^{d-1} 2 E_i}\right)
(2\pi)^d \mu^{-2\ve}
\delta^d\left(p_A^\mu+p_B^\mu-\sum_{i=1}^n p_i^\mu\right)
\eeq
($n=N,N+1$) we included the identical particle factor $1/n!$ which is 
present if we treat all final state partons identical and sum over all
possible parton flavors $a_i=u,\bar{u},d,\bar{d},\ldots,g$. The parton
distribution functions for incoming partons $A$ and $B$ defined in the
\ms\ renormalization scheme are denoted by $f_A(a_A,x_A)$ and
$f_B(a_B,x_B)$.

In order to factor the dependence of
the cross section on the physics of low transverse momenta out of the
partonic cross section and into these \ms\ parton distributions, in the
$2\to N$ cross section one uses the modified parton distribution
$\f(a,x)$ that satisfies
\beqn
\label{ftilde}
&&\f(a,x)=\sum_b \int\frac{\d z}{z}f(b,x/z)
\\ \nn && \qquad\qquad
\times\left[\delta_{ab}\delta(1-z)
+\frac{(4\pi)^\ve}{\ve\Gamma(1-\ve)}
\frac{\as}{2\pi}P_{a/b}(z)+\O(\as^2)\right],
\eeqn
with $P_{a/b}(z)$ being the full \AP kernel, for the $b\to a$ splitting:
\beq
\label{APkernel}
P_{a/b}(z)=\P_{a/b}(z)-\delta_{ab}\frac{2C(a)}{1-z}
+\delta_{ab}\frac{2C(a)}{(1-z)_+}+\delta_{ab}\gamma(a)\delta(1-z),
\eeq
where, for instance, in the case of $g\to gg$ splitting
\beq
\label{APsplit}
\P_{g/g}(z) = 2C(g)\left(\frac{1-z}{z}+\frac{z}{1-z}+z(1-z)\right).
\eeq
$C(g)=\Nc$ is the color charge of the gluon, while $C(q)=V/(2\Nc)$
is that of the quark ($V=(\Nc^2-1)$), while the $\gamma$ constants
represent the contribution of the virtual graphs to the \AP kernel,
\beq
\gamma(g)=\frac{11\Nc-2N_f}{6},\quad
\gamma(q)=\frac{3V}{4\Nc}.
\eeq
The notation $1/(1-z)_+$ is the usual ``+'' prescription,
\beq
\int_0^1 \frac{f(z)}{(1-z)_+} =
\int_0^1 \frac{f(z)-f(1)}{1-z}.
\eeq
This factorization recipe is discussed in ref.\ \cite{mueller}. The
$\la |\M (2\to n)|^2 \ra$ functions are the $2\to n$ squared matrix elements 
averaged over initial state and summed over final state spins and colors:
\beq
\la |\M (a+b\to n)|^2 \ra = \frac{1}{\omega(a) \omega(b)}
\sum_{{\rm spin} \choose {\rm color}} |\M (a+b\to n)|^2.
\eeq
In the conventional \ms\ scheme, we need $\la |\M|^2 \ra$ in $d=4-2\ve$
dimensions. However, it was shown in ref.\ \cite{KST2to2} that simple 
rules exist which tell us how to obtain the finite $2\to N$ hard
scattering cross section of the conventional \ms\ scheme at \NLO 
using the expressions for $\la |\M (2\to n)|^2 \ra$ obtained  in the
dimensional reduction scheme. Therefore, we use $\omega(g)=2V$ and
$\omega(q)=2\Nc$, which are valid in $d=4$ dimensions 
and the four-dimensional expressions for the squared matrix elements.
Finally, the functions $\S_n$ define the physical quantity to be
calculated.

In equation (\ref{physical}) both terms are singular when the
regularization is removed, $\ve \to 0$. When $\ve \ne 0$ the singularities
are represented as $1/\ve^2$ and $1/\ve$ poles. These poles cancel
between the $\I[2\to N]$ and $\I[2\to N+1]$ terms, provided the physical
measurement, represented by the functions $\S_n$, is infrared safe.
This means that the emission of a soft or a collinear parton must not
influence the result of the measurement. Therefore, the measurement
functions $\S_n$ must possess the following properties:
\beq
\label{softcond}
\lim_{p_i^\mu\to 0} \S_{N+1}(p_1^\mu,\ldots,p_{N+1}^\mu)
= \S_N(p_1^\mu,\ldots,\del{p_i^\mu},\ldots,p_{N+1}^\mu), 
\quad i\in [1,N+1];
\eeq
\beqn
\label{collcond}
&&\lim_{{p_i^\mu \to z p_P^\mu} \choose 
      {p_j^\mu \to (1-z) p_P^\mu}}\S_{N+1}(p_1^\mu,\ldots,p_{N+1}^\mu)
= \S_N(p_1^\mu,\ldots,\del{p_i^\mu},p_P^\mu,
\ldots,\del{p_j^\mu},\ldots,p_{N+1}^\mu),
\nn \\ \nn &&
\qquad \qquad i,j\in [1,N+1], \quad z\in [0,1],
\\ && ~ \\ 
&&\lim_{p_j^\mu \to (1-z) p_m^\mu}\S_{N+1}(p_1^\mu,\ldots,p_{N+1}^\mu)
= \S_N(p_1^\mu,\ldots,\del{p_j^\mu},\ldots,p_{N+1}^\mu),
\nn \\ \nn &&
\qquad \qquad m=A,B,\quad j\in [1,N+1], \quad z\in [0,1].
\eeqn

\section{The $2\to N$ integral}
\setcounter{equation}{0}

In this section, our aim is to write the $\I[2\to N]$ integral in such a
form that will make the cancelation of divergent pieces against 
corresponding divergent terms in the $\I[2\to N+1]$ integral as simple
as possible. The discussion is a generalization of the corresponding 
discussion in ref.\ \cite{KSjets} given for the $\I[2\to 2]$ integral 
to the $2\to N$ case. There are some differences however. Firstly, we 
do not specify the integration variables, but leave it for the reader 
to use a preferred choice. Secondly, in order we could use the results 
for the five-parton one-loop QCD helicity amplitudes
\cite{BDK5g,KST4q1g,BDK2q3g} for a \NLO calculation of three-jet
production in hadron collisions, we perform the analysis
using dimensional reduction scheme and add the necessary transition terms
at the end to obtain the correct formula in conventional \ms\ scheme
as they are given in ref.\ \cite{KST2to2}.

In order to simplify the book-keeping of the various factors, we introduce
the integration measure
\beq
\label{Ddef}
\D_N(\ve)= 
\frac{1}{2s}\frac{1}{N!}\frac{\as^N}{(2\pi)^{2N-4}}\, \d x_A\,\d x_B\,
\prod_{i=1}^N\left[(2\pi\mu)^{2\ve}\d^{4-2\ve}p_i 2 \delta(p_i^2)\right],
\eeq
where the subscript on $\D$ reminds us that this measure is related
to the phase space integration measure of $N$ particles according to the
relation
\beq
\label{ps2toN}
\frac{(4\pi\as)^N}{2s}
\d x_A\,\d x_B\,\d\Gamma^{(N)}(p_1^\mu,\ldots,p_N^\mu) = \D_N(\ve) 
(2\pi\mu)^{-2\ve}\delta^d\left(p_A^\mu+p_B^\mu-\sum_{i=1}^N p_i^\mu\right).
\eeq

It is convenient to write the perturbative expansion of the squared
matrix element summed over final spins and colors and averaged over
initial spins and colors and with ultraviolet renormalization in the 
\ms\ renormalization scheme included in terms of functions 
$\Psi^{(2N)}_{{\rm DR}}(\vec{p})$ and
$\Psi^{(2N+2)}_{{\rm DR}}(\vec{p})$,
\beqn
\label{nlom2}
\lefteqn{\la |\M(2\to N)|^2\ra } \\ \nn
&=& \frac{(4\pi\as)^N}{\omega(a_A)\omega(a_B)}
\left\{\Psi^{(2N)}_{{\rm DR}}(\vec{a},\vec{p})
+\frac{\as}{2\pi}\cG\left(\frac{\mu^2}{\qes}\right)^\ve
\Psi^{(2N+2)}_{{\rm DR}}(\vec{a},\vec{p})
+{\rm O}\left(\left(\frac{\as}{2\pi}\right)^2\right)\right\},
\eeqn
where
\beq
\cG=\left(4\pi\right)^\ve
\frac{\Gamma^2(1-\ve)\Gamma(1+\ve)}{\Gamma(1-2\ve)},
\eeq
and the subscript DR refers to expressions obtained using dimensional
reduction. We use the notation $\vec{a} = (a_A,a_B, a_1,\ldots,a_N)$
for the collection of external parton flavors and $\vec{p} =
(p^\mu_A,p^\mu_B, p^\mu_1,\ldots,p^\mu_N)$ for the collection of
external four-momenta.
The variable $Q_{\rm ES}$ is an arbitrary parameter of mass dimensions
introduced to facilitate writing the result \cite{ES}. The dependence
of the function $\Psi^{(2N+2)}_{\rm DR}$ on $Q_{\rm ES}$ is such that
the squared matrix element does not actually depend on $Q_{\rm ES}$.
The first term in the curly braces is the Born $2\to N$ matrix element
squared, without the $g^{2N}$ coupling factor, while the second term
is the \NLO contribution, which in the dimensional reduction scheme can 
be expressed in terms of the helicity amplitudes \cite{KST2to2},
\beq
(4\pi\as)^N
\frac{\as}{2\pi}\cG\left(\frac{\mu^2}{\qes}\right)^\ve
\Psi^{(2N+2)}_{{\rm DR}}(\vec{a},\vec{p})
=\sum_{{\rm hel}} \sum_{{\rm col}}
\left(\Am^{(1)}\Am^{(0)*}+\Am^{(1)*}\Am^{(0)}\right).
\eeq
On the right hand side the superscript (0) and (1) refers to the tree
and one-loop helicity amplitudes respectively. The helicity amplitudes
can be decomposed in color space in terms of gauge invariant color
subamplitudes. For instance, in the case of pure gluon processes,
\beq
\Am^{(i)}(g_1,\ldots,g_N) =
g^N \left(\frac{g}{4\pi}\right)^{(2i)}
\sum_n C_n^{g_1,\ldots,g_N}
a^{(i)}_n(1,\ldots,N),
\eeq
where $C^{g_1,\ldots,g_N}$ is a general invariant matrix in color space 
with upper indices in the adjoint representation and summation on $n$ runs 
over a linearly independent set of such matrices. Explicit examples of 
this decomposition can be found in reference \cite{BKcolor}.
The one-loop color subamplitudes and thus the helicity amplitudes can 
naturally be decomposed into singular terms containing at most double
poles in $\ve$ and into terms that are finite when $\ve\to 0$,
\beq
\Am^{(1)}=\Am^{(1)}_{{\rm S}}+\Am^{(1)}_{{\rm NS}}.
\eeq
Looking at the explicit form of the singular terms of one-loop five-parton
color subamplitudes \cite{KSTsing}, we see that the imaginary parts of the
factors $-1/\ve^2(-\s ij/\qes)^{-\ve}$ do not contribute to the function
$\Psi^{(8)}_{\rm DR}$. In such cases the $\Psi^{(2N+2)}_{\rm DR}$
functions have the following structure:
\beqn
\label{Psi2N+2structure}
\Psi^{(2N+2)}_{{\rm DR}}(\vec{a},\vec{p})&=&
\Psi^{(2N)}_{{\rm DR}}(\vec{a},\vec{p})
\left\{-\frac{1}{\ve^2}\sum_{n=A,B,1,\ldots,N} C(a_n)
-\frac{1}{\ve}\sum_{n=A,B,1,\ldots,N} \gamma(a_n)\right\} \\ \nn
&& +\frac{1}{2\ve}\sum_{{m,n=A,B,1,\ldots,N} \choose {m\ne n}}^N\ell(\s mn)
\Psi^{(2N;c)}_{mn,\,{\rm DR}}(\vec{a},\vec{p}) \\ \nn
&& + \Psi^{(2N)}_{{\rm DR}}(\vec{a},\vec{p})\ell(\mu^2)
\sum_{n=A,B,1,\ldots,N} \gamma(a_n)
-\frac{1}{4}\sum_{{m,n=A,B,1,\ldots,N} \choose {m\ne n}}^N\ell_2(\s mn)
\Psi^{(2N;c)}_{mn,\,{\rm DR}}(\vec{a},\vec{p}) \\ \nn
&& + 2 \left[ (4\pi\as)^N
\frac{\as}{2\pi}\cG\left(\frac{\mu^2}{\qes}\right)^\ve\right]^{-1}
\frac{1}{2} \sum_{{\rm hel}} \sum_{{\rm col}}
\left(\Am^{(1)}_{{\rm NS}}\Am^{(0)*}
+\Am^{(1)*}_{{\rm NS}}\Am^{(0)}\right)+{\rm O}(\ve).
\eeqn
Here the $\Psi^{(2N;c)}_{mn}$ functions are the color correlated
Born squared matrix elements defined in ref.\ \cite{KSjets}.
The factor $2 \left[ (4\pi\as)^N (\as/2\pi)\cG
\left(\mu^2/\qes\right)^\ve\right]^{-1}=
\left[g^{2N}\left(g/4\pi\right)^2\right]^{-1} + {\rm O}(\ve)$
cancels against the coupling factors in $\Am^{(1)}_{{\rm NS}}\Am^{(0)}$.
The functions $\ell(x)$ and $\ell_2(x)$ are defined as
\beq
\label{elldef}
\ell(x) = \ln\left|\frac{x}{\qes}\right|,\qquad
\ell_2(x)=\ell^2(x)-\pi^2\Theta(x).
\eeq

Substituting the integration measure of eq.\ (\ref{ps2toN}), 
the perturbative expression for the squared matrix element,
eq.\ (\ref{nlom2}) and the expression for the modified effective parton
distribution functions as defined in eq.\ (\ref{ftilde}), we can write 
the $2\to N$ cross section as
\beqn
\label{I2toN}
&&\I[2\to N]_{\rm DR} =
\\ \nn &&
\sum_{a_A,a_B,a_1,\ldots,a_N}\int \D_N(\ve)\S_N(p_1^\mu,\ldots,p_N^\mu)
(2\pi\mu)^{-2\ve}\delta^d\left(p_A^\mu+p_B^\mu-\sum_{i=1}^N p_i^\mu\right)
\\ \nn &&\qquad\qquad\quad
\times\Bigg\{L(a_A,a_B,x_A,x_B)
\left[\Psi^{(2N)}_{{\rm DR}}(\vec{a},\vec{p})
+\frac{\as}{2\pi}\cG\left(\frac{\mu^2}{\qes}\right)^\ve
\Psi^{(2N+2)}_{{\rm DR}}(\vec{a},\vec{p})\right]
\\ \nn &&\qquad\qquad\qquad
+\sum_{a_A'}\frac{\omega(a_A')}{\omega(a_A)}\int_{x_A}^1\frac{\d z}{z^2}
L\left(a_A',a_B,\frac{x_A}{z},x_B\right)\frac{(4\pi)^\ve}{\ve\Gamma(1-\ve)}
\frac{\as}{2\pi}P_{a_A/a_A'}(z)\Psi^{(2N)}_{{\rm DR}}(\vec{a},\vec{p})
\\ \nn &&\qquad\qquad\qquad
+\sum_{a_B'}\frac{\omega(a_B')}{\omega(a_B)}\int_{x_B}^1\frac{\d z}{z^2}
L\left(a_A,a_B',x_A,\frac{x_B}{z}\right)\frac{(4\pi)^\ve}{\ve\Gamma(1-\ve)}
\frac{\as}{2\pi}P_{a_B/a_B'}(z)\Psi^{(2N)}_{{\rm DR}}(\vec{a},\vec{p}) 
\Bigg\}.
\eeqn
The function $L$ used here and elsewhere describes the parton
luminosity:
\beq
L(a_A,a_B,x_A,x_B) = 
\frac{f(a_A,x_A)}{\omega(a_A)x_A}
\frac{f(a_B,x_B)}{\omega(a_B)x_B}.
\eeq

According to the factorization and Kinoshita-Lee-Nauenberg theorems, and 
we shall see it at \NLO explicitly in the following sections, the pole 
terms in eq.\ (\ref{I2toN}) cancel against poles emerging in the phase 
space integral of the brehmsstrahlung contributions. Therefore, it is 
only the Born function $\Psi^{(2N)}$ and the finite part of the
$\Psi^{(2N+2)}$ function --- the last two lines of eq.\ 
(\ref{Psi2N+2structure}) --- that is really integrated in eq.\ 
(\ref{I2toN}). However, we need the corresponding finite expressions
valid in conventional dimensional regularization. In ref.\ \cite{KST2to2}
it was shown that simple terms are to be added in order to obtain the
correct formula we need for a \NLO calculation in the conventional \ms\
scheme. Thus the function resulting from the loop corrections
that we need for a \NLO Monte Carlo program is the non-singular function
\beqn
\label{Psi2N+2ns}
\Psi^{(2N+2)}_{{\rm NS}}(\vec{a},\vec{p})&=&
\Psi^{(2N)}_{{\rm DR}}(\vec{a},\vec{p})
\left[\sum_{n=A,B,1,\ldots,N}\left[\ell(\mu^2)\gamma(a_n) 
- \tilde{\gamma}(a_n)\right] + N\frac{\Nc}{6}\right] \\ \nn
&& -\frac{1}{4}\sum_{{m,n=A,B,1,\ldots,N} \choose {m\ne n}}\ell_2(\s mn)
\Psi^{(2N;c)}_{mn,\,{\rm DR}}(\vec{a},\vec{p}) \\ \nn
&& + \left[g^{2N}\left(\frac{g}{4\pi}\right)^2\right]^{-1}
\frac{1}{2} \sum_{{\rm hel}} \sum_{{\rm col}}
\left(\Am^{(1)}_{{\rm NS}}\Am^{(0)*}
+\Am^{(1)*}_{{\rm NS}}\Am^{(0)}\right),
\eeqn
where the transition terms $\tilde{\gamma}(a_n)$ are given by
\beq
\tilde{\gamma}(g) = \frac{1}{6}C(g),\quad
\tilde{\gamma}(q) = \frac{1}{2}C(q).
\eeq

\section{The $2\to N+1$ integral}
\setcounter{equation}{0}

In this section, we separate the $\I[2\to N+1]$ integral
into terms containing $1/\ve^2$ and $1/\ve$ poles, which cancel against
the corresponding poles of the $\I[2\to N]$ integral, and terms that
are finite when $\ve \to 0$ and, therefore, can be integrated
numerically. The $1/\ve^p$ singularities arise from integrating the
square of the matrix element over the $(N+1)$-particle phase space when
a gluon becomes soft, or two partons become collinear. Firstly, we 
organize the integration domain  so as to reduce the complexity of the
problem.

\subsection{The domain of integration}

We must integrate over the momenta of the $N+1$ final state particles
treating them identical. If we do not fix a definite label to each 
particle, then we integrate over each event topology $(N+1)!$ times. We 
can however, simplify the calculation by \begin{enumerate}
\item first splitting the phase space {\em in the parton-parton c.m.
system} into two parts: in the first one, the smallest angular distance 
$r_{ij}=\s ij/(E_i E_j)$ is between final state particles,
while in the second region the smallest angular distance is between an
initial state particle and a final state particle;
\item secondly cutting into the first region by fixing that label of the 
smallest Lorentz invariant of final state particle pairs to which the 
smaller energy {\em in the parton-parton c.m. system} belongs to be 
$j=N+1$, and cutting into the second region by fixing the final state 
label of the smallest Lorentz invariant of pairs involving an initial
state and a final state particle to be $j=N+1$.
\end{enumerate}
With this distinction of parton $(N+1)$, we have to integrate 
over each event topology only $N!$ times and there is a corresponding 
symmetry factor $1/N!$ associated with the integration:
\beqn
\label{psN+1mod}
&&\d \Gamma^{(N+1)}(p_1^\mu,\ldots,p_{N+1}^\mu) =
\frac{1}{N!} \prod_{i=1}^{N+1} 
\left(\frac{\mu^{2\ve}\d^{d-1}{\bf p}_i}{(2\pi)^{d-1} 2 E_i}\right)
(2\pi)^d \mu^{-2\ve}
\delta^d\left(p_A^\mu+p_B^\mu-\sum_{i=1}^{N+1} p_i^\mu\right)
\\ \nn &&\qquad\qquad\qquad\qquad\qquad \times
\big[\Theta(r^{(N+1;i)}_{\rm min} > r^{(N+1;f)}_{\rm min})
\Theta(s^{(N;f)}_{\rm min} > \s k{,N+1}^{\rm min}) \Theta(E_k > E_{N+1})
\\ \nn &&\qquad\qquad\qquad\qquad\qquad\quad 
+\Theta(r^{(N+1;f)}_{\rm min} > r^{(N+1;i)}_{\rm min})
\Theta(s^{(N;i)}_{\rm min} > \s X{,N+1}^{\rm min})\big],
\eeqn
where 
\beqn
&&d^{(n;f)}_{\rm min} = \min (d_{ij} : i,j=1,\ldots,n,\; i\ne j),\\
&&d^{(n;i)}_{\rm min} = \min (d_{Xj} : X=A,B,\;j=1,\ldots,n),\\
\label{dkmin}
&&d_{k,N+1}^{\rm min} = \min (d_{j,N+1} : j=1,\ldots,N),\\
&&d_{X,N+1}^{\rm min} = \min (d_{A,N+1},\;d_{B,N+1}),
\eeqn
with $d$ meaning either Lorentz invariant $s$ or angular distance $r$.
In eq.\ (\ref{dkmin}) the index $k$ denotes that $j$ for which the
minimum value is assumed. 

In the cut phase space the only singularities that can occur when a
single Lorentz invariant vanishes are
\begin{itemize}
\item parton $N+1$ is soft (in both regions);
\item parton $N+1$ is collinear to a final state parton $i\in [1,N]$
(in the first region);
\item parton $N+1$ is collinear to an initial state parton A or B
(in the second region).
\end{itemize}
We are not interested in configurations when two Lorentz invariants
involving four different labels vanish simultaneously, because those
emerge only in $(N\!-\!1)$-jet configurations that are not considered here.

Next, we study the singularity structure of the squared matrix element.

\subsection{Singularity structure of the squared matrix element}

In order to find the singularity structure of $\I[2\to N+1]$ over the cut
phase space of eq.\ (\ref{psN+1mod}) explicitly, it is 
useful to strip off the spin and color averaging and the coupling of 
the squared matrix element. We define the function
\beq
\Psi^{(2N+2)} (A,B,1,\ldots,N+1) =
\frac{\omega(a_A)\omega(a_B)}{(4\pi\as)^{(N+1)}}
\la |\M (2\to N+1)|^2 \ra ,
\eeq
where the argument denotes:
\beq
(A,B,1,\ldots,N+1)\equiv
(a_A, a_B, a_1,\ldots a_{N+1};p^\mu_A, p^\mu_B, p^\mu_1,\ldots p^\mu_{N+1}).
\eeq
The singularity structure of the function $\Psi^{(2N+2)}$ in four dimensions 
can most easily be obtained from the factorization properties of helicity
amplitudes \cite{MP90}. Citing only the
results, we find for soft gluon labeled $j=N+1$
\beq
\label{soft}
\lim_{p_j^\mu\to 0} \Psi^{(2N+2)}(A,B,1,\ldots,N+1) = 
\sum_{{m,n = A,B,1,\ldots N} \choose {m<n}}
\delta_{a_jg}\frac{2 \s mn}{\s mj \s jn} \Psi^{(2N;c)}_{mn}(\vec{a},\vec{p})
+\O\left(\frac{1}{\sqrt{\s mj}},\frac{1}{\sqrt{\s nj}}\right),
\eeq
which is called ``soft identity'' in ref.\ \cite{KSjets}. 
In order to make the cancelation of the infrared singularities as
transparent as possible, it is useful to perform single singular 
decomposition of the eikonal factor in eq.\ (\ref{soft}):
\beq
\frac{2\s mn}{\s mj \s jn} =
 \frac{2\s mn}{\s mj (\s mj + \s nj)}
+\frac{2\s mn}{\s nj (\s mj + \s nj)}.
\eeq
With this decomposition and using the symmetry of the 
$\Psi^{(2N;c)}_{mn}$ functions in the $m$, $n$ indices we can write eq.\
(\ref{soft}) in the form
\beq
\label{softdecomp}
\lim_{p_j^\mu\to 0} \Psi^{(2N+2)}(A,B,1,\ldots,N+1) = 
\sum_{{m,n = A,B,1,\ldots N} \choose {m\ne n}}
\delta_{a_jg} \Psi_{S;mn}^{(2N+2)}(\vec{a},\vec{p},p_j^\mu),
\eeq
where
\beq
\label{PsiSmn}
\Psi_{S;mn}^{(2N+2)}(\vec{a},\vec{p},p_j^\mu) = 
\frac{2\s mn}{\s mj (\s mj + \s nj)}
\Psi^{(2N;c)}_{mn}(\vec{a},\vec{p}).
\eeq

In the collinear limit of two final state partons $i$ and $j=N+1$,
we introduce a pseudo particle $P$ with $a_P$ flavor 
that splits into gluons $i$ and $j$: $p^\mu_P=p^\mu_i+p^\mu_j$. The
flavor $a_P=a_i$ if $a_j=g$ and $a_P=g$ if $a_i=q$, $a_j=\bar{q}$.
The momentum fraction $z$ is defined by $p^\mu_i=z p^\mu_P$. 
Then for the collinear limit of $\Psi^{(2N+2)}$ one finds:
\beqn
&&\lim_{{p_i^\mu \to z p_P^\mu} \choose 
      {p_j^\mu \to (1-z) p_P^\mu}} \Psi^{(2N+2)}(A,B,1,\ldots,N+1)
\\ \nn &&\qquad\qquad\qquad
= \frac{2}{\s ij} \Psi_{C;ij}^{(2N+2)}(z,P;A,B,1,\ldots,N+1)
+\O\left(1/\sqrt{\s ij}\right),
\eeqn
where
\beqn
\label{collfin}
&&\Psi_{C;ij}^{(2N+2)}(z,P;A,B,1,\ldots,N+1)
\\ \nn &&\qquad\qquad
=\P_{a_i/a_P}(z) \Psi^{(2N)}(A,B,\ldots,\del i,P,\ldots,\del j,\ldots) 
\\ \nn &&\qquad\qquad\quad
+\,2 \Re e \left(Q_{P\to ij}(z)
\Phi^{(2N)}(P;A,B,\ldots,\del{i},\ldots,\del{j},\ldots)\right).
\eeqn
In this equation $\Psi^{(2N)}(A,B,\ldots,\del i,P,\ldots,\del j,\ldots)$
is the $\Psi^{(2N)}$ function of $2+N$ partons obtained from 
$\Psi^{(2N+2)}(2\to N+1)$ by deleting labels $i$ and $j$ and adding 
the pseudo particle label $P$, $\P_{a_i/a_P}(z)$ is the \AP splitting
function for the process $P\to ij$ in four dimensions without $z=1$
regulation (eq.\ (\ref{APsplit}) in the case of gluon splitting).
The $Q_{P\to ij}(z)$ functions are calculated from the tree-level
splitting amplitudes, $\Sp_\lambda(i^{h_i},j^{h_j})$ of
ref.\ \cite{BDKcoll} according to the formula
\beq
\frac{2}{\s ij}Q_{P\to ij}(z) = \sum_{i,j}\sum_{h_i,h_j=\pm}
c(i,j,P) c(i,j,P)^* \Sp_-(i^{h_i},j^{h_j})\Sp_+(i^{h_i},j^{h_j})^*
\eeq
where $c(i,j,P)$ is the color matrix of the $P\to ij$ vertex.
In the case of gluon splitting
\beq
Q_{g\to g_i g_j}(z) =-2C(g)z(1-z)\frac{\A ij}{\B ij},\quad
Q_{g\to q_i \bar{q}_j}(z) = z(1-z)\frac{\A ij}{\B ij},
\eeq 
while in the case of quark splitting $Q_{q\to ij}(z)=0$, which is
also understood from helicity conservation along fermion lines.
The function $\Phi^{(2N)}$ does not depend on the momenta $p_i^\mu$ 
and $p_j^\mu$ only on their sum, $p_P^\mu$. The $\Phi^{(2N)}$
functions are calculated from the tree-level helicity amplitudes as the
Born function $\Psi^{(2N)}$, except that the summation over the helicity
of parton $P$ is not carried out:
\beqn
&&\Phi^{(2N)}(P;A,B,\ldots,\del{i},\ldots,\del{j},\ldots)
\\ \nn &&\qquad
=\sum_{\rm color}\sum_{h_A,h_B,\ldots}
\Am^{(0)}(A^{h_A},B^{h_B},\ldots,\del{i},P^+,\ldots,\del{j},\ldots)
\\ \nn &&\qquad\qquad\qquad\qquad\times
\Am^{(0)}(A^{h_A},B^{h_B},\ldots,\del{i},P^-,\ldots,\del{j},\ldots)^*
\eeqn

In the collinear limit of a final state parton $j=N+1$ with an initial
state parton $A$, we let $A$  split into partons $P$ and $j$: 
$p^\mu_A=p^\mu_P+p^\mu_j$, with momentum fraction $z$ defined as 
$p^\mu_P=z p^\mu_A$, followed by an $P+B\to 1,\ldots N$ hard-scattering 
process. From the crossing of $A$ and $P$ in relation (\ref{collfin}),
for the collinear limit of $\Psi^{(2N+2)}$ one obtains:
\beqn
&&\lim_{p_j^\mu \to (1-z) p_A^\mu} \Psi^{(2N+2)}(A,B,1,\ldots,N+1)
\\ \nn &&\qquad\qquad
=-\frac{2}{\s Aj}(-1)^{f(a_A)+f(a_P)}
\Psi_{C;\bar{A}j}^{(2N+2)}(1/z,P;A,B,1,\ldots,N+1)
+\O\left(1/\sqrt{\s Aj}\right),
\eeqn
where 
\beq
f(g)=0,\quad f(q)=1.
\eeq
We can write the right hand side of eq.\ (\ref{collini}) in a more
explicit form using the crossing relation of the \AP splitting functions,
\beq
\P_{b/a}(z)=-(-1)^{f(a)+f(b)}\frac{\omega(b)}{\omega(a)}
z\P_{\bar{a}/\bar{b}}(1/z),
\eeq
where $\bar{a}$ is the antiparticle of particle $a$. Thus we find
\beqn
\label{collini}
&&-(-1)^{f(a_A)+f(a_P)}
\Psi_{C;\bar{A}j}^{(2N+2)}(1/z,P;A,B,1,\ldots,N+1)
\\ \nn &&\qquad\qquad\qquad
=\frac{\omega(a_A)}{\omega(a_P)}\frac{1}{z}\P_{a_P/a_A}(z)
\Psi^{(2N)}(P,B,\ldots,\del j,\ldots) 
\\ \nn &&\qquad\qquad\qquad\quad
-(-1)^{f(a_A)+f(a_P)}
2\Re e\left(Q_{\bar{P}\to \bar{A}j}(1/z)
\Phi^{(2N)}(P;B,\ldots,\del j,\ldots)\right).
\eeqn

We close the analysis of the singularity structure of $\Psi^{(2N+2)}$
with considering the limit when the soft gluon is also collinear with 
parton $m$. Thus we take $p_j^\mu=(1-z)p_m^\mu$ with $z\to1$. Using
the ``soft-collinear'' identity of ref.\ \cite{KSjets},
\beq
\label{softcollidentity}
\sum_{{n=A,B,1,\ldots,N} \choose {n\ne m}}
\Psi^{(2N;c)}_{mn}(\vec{p})=
2C(a_m)\Psi^{(2N)}(A,B,1,\ldots,N),
\eeq
we obtain from eq.\ (\ref{PsiSmn})
\beq
\label{softcoll}
\lim_{p_j^\mu\to(1-z)p_m^\mu}
\sum_{{n=A,B,1,\ldots,N} \choose {n\ne m}}
\delta_{a_jg}\Psi^{(2N+2)}_{S;mn}(\vec{p})=
2C(a_m)\delta_{a_jg}\frac{2}{(1-z)\s mj}\Psi^{(2N)}(\vec{p}).
\eeq
In the same limit, eqs.\ (\ref{collfin}) and (\ref{collini}) yield
\beq
\label{softcoll2}
\lim_{z\to 1}
\frac{2}{\s mj} \Psi^{(2N+2)}_{C;mj}(z,P;A,B,1,\ldots,N+1)=
2C(a_m)\delta_{a_jg}\frac{2}{(1-z)\s mj}\Psi^{(2N)}(\vec{p})
\eeq
and
\beq
\label{softcoll3}
\lim_{z\to 1}
\frac{2}{\s Aj} \Psi^{(2N+2)}_{C;Aj}(1/z,P;A,B,1,\ldots,N+1)=
2C(a_A)\delta_{a_jg}\frac{2}{(1-z)\s Aj}\Psi^{(2N)}(\vec{p}),
\eeq
in agreement with eq.\ (\ref{softcoll}).

The function $\Psi^{(2N+2)}$ does not posses any other poles when 
parton $N+1$ is soft or collinear to another parton. Knowing the 
singularity structure of the function $\Psi^{(2N+2)}$, we can decompose 
the integral $\I[2\to N+1]$ into three terms:
\beq
\label{2toN+1decomp}
\I[2\to N+1] = \I[\soft]+\I[\coll]+\I[\fin].
\eeq
The first two of these integrals possess divergences in $\ve$, 
but they are sufficiently simple to calculate the pole structure 
of the Laurent expansion in $\ve$ around zero analytically.
The third one is complicated, but contains at most square-root 
singularities over the modified $(N+1)$-particle phase space of eq.\
(\ref{psN+1mod}), therefore, can be integrated numerically yielding a 
finite contribution as $\ve\to 0$. We further decompose the ``soft''
and ``collinear'' contributions into sums of $N+2$ terms,
\beqn
&&\I[\soft] = \sum_{m=A,B,1,\ldots,N} \I[\soft]_m, \\
&&\I[\coll] = \sum_{m=A,B,1,\ldots,N} \I[\coll]_m,
\eeqn
where $\I[\soft]_m$ is associated with the integral of the soft terms
$\Psi^{(2N+2)}_{S;mn}$, while $\I[\coll]_m$ is associated with the
integral of the collinear term $\Psi^{(2N+2)}_{C;m,N+1}$. We shall
call these integrals subtraction terms for the obvious reason that
subtracting them from the $\I[2\to N+1]$ integral the finite term
remains. In order to define the soft and collinear subtraction terms
precisely, we first give a decomposition of the phase space into 
such regions that in any one of them only one pair of external
particles can become collinear.

\subsection{Decomposition of the phase space integral}

In this subsection, our goal is to write the phase space integral in those
variables that allow the most efficient Monte Carlo integration.
We write the integration measure of eq.\ (\ref{psN+1mod}) as
\beq
\label{psdecomp}
\frac{(4\pi\as)^{(N+1)}}{2s}
\d x_A\,\d x_B\,\d\Gamma^{(N+1)}(p_1^\mu,\ldots,p_{N+1}^\mu) =
\frac{\as}{(2\pi)^2}
\Bigg[\sum_{m_f=1}^N\D_{N+1}^{\fin;m_f}(\ve)
+\sum_{m_i=A,B}\D_{N+1}^{\ini;m_i}(\ve)\Bigg],
\eeq
where 
\beqn
&&\D_{N+1}^{\fin;m}(\ve)=
\D_N(\ve)\,
(2\pi\mu)^{2\ve} \d^{4-2\ve} p\,2\,\delta\!\left(p^\mu p_\mu\right)\,
(2\pi\mu)^{-2\ve}
\delta^d\left(p_A^\mu+p_B^\mu-\sum_{i=1}^{N+1} p_i^\mu\right)
\\ \nn &&\qquad\qquad\quad\times
\Theta(r^{(N+1;i)}_{\rm min} > r^{(N+1;f)}_{\rm min})
\Theta(s^{(N;f)}_{\rm min} > \s k{,N+1}^{\rm min}) \Theta(E_k > E_{N+1})
\\ \nn &&\qquad\qquad\quad\times
\Theta(\min(r_{i,N+1}: i=1,\ldots,N, ~ i\ne m) > r_{m,N+1}),
\eeqn
\beqn
&&\D_{N+1}^{\ini;A}(\ve)=
\D_N(\ve)\,
(2\pi\mu)^{2\ve} \d^{4-2\ve} p\,2\,\delta\!\left(p^\mu p_\mu\right)\,
(2\pi\mu)^{-2\ve}
\delta^d\left(p_A^\mu+p_B^\mu-\sum_{i=1}^{N+1} p_i^\mu\right)
\\ \nn &&\qquad\qquad\quad\times
\Theta(r^{(N+1;f)}_{\rm min} > r^{(N+1;i)}_{\rm min})
\Theta(s^{(N;i)}_{\rm min} > s_{X,N+1}^{\rm min})
\Theta(r_{B,N+1} > r_{A,N+1}),
\eeqn
with $p^\mu \equiv p_{N+1}^\mu$ and $\D_N(\ve)$ is defined in equation 
(\ref{Ddef}). The measure $\D_{N+1}^{\ini;B}$ is defined as 
$\D_{N+1}^{\ini;A}$ with $A$ and $B$ interchanged. The advantage of this 
decomposition of the phase space should be clear: in each region there is
only one pair of particles that can become collinear. As a result, the
single singular factor decomposition of the integrand is substituted by a 
(much simpler) decomposition of the phase space.

In order to write the integration measure $\D_{N+1}^{\fin;m}$ in the 
required form, we utilize a vector $p_Q^\mu$ of invariant mass $Q^2$ that 
splits into the vectors $p_m^\mu$ and $p^\mu\equiv p_{N+1}^\mu$ and use 
the mathematical identity
\beqn
\label{identity}
&&\D_{N+1}^{\fin;m}(\ve) =
\int_0^{\hat{s}}\frac{\d Q^2}{2\pi}\Bigg[\D_N(\ve)\,(2\pi\mu)^{-2\ve}
\delta^d\left(p_A^\mu+p_B^\mu-\sum_{i=1}^{N} p_i^\mu\right)
\Bigg]\st{m\to Q}
\\ \nn &&\qquad\qquad\qquad\qquad\times
\frac{\mu^{2\ve}\d^{d-1}{\bf p}_m}{(2\pi)^{d-1} 2 E_m}
(2\pi\mu)^{2\ve}\frac{\d^{d-1}{\bf p}}{E}
(2\pi)^d \mu^{-2\ve}
\delta^d\left(p_Q^\mu-p_m^\mu-p^\mu\right)
\\ \nn &&\qquad\qquad\qquad\qquad\times
\Theta(r^{(N+1;i)}_{\rm min} > r^{(N+1;f)}_{\rm min})
\Theta(s^{(N;f)}_{\rm min} > \s k{,N+1}^{\rm min}) \Theta(E_k > E)
\\ \nn &&\qquad\qquad\qquad\qquad\times
\Theta(\min(r_{i,N+1}: i=1,\ldots,N, ~ i\ne m) > r_{m,N+1}).
\eeqn
We can use the second $\delta$ function for integrating over the $d-1$ 
momenta of particle $m$ and over $Q^2$. We obtain
\beq
\label{pmintegral}
\int_0^{\hat{s}} \frac{\d Q^2}{2\pi}
\frac{\mu^{2\ve}\d^{d-1}{\bf p}_m}{(2\pi)^{d-1} 2 E_m}
(2\pi\mu)^{2\ve}\frac{\d^{d-1}{\bf p}}{E}
(2\pi)^d \mu^{-2\ve}
\delta^d\left(p_Q^\mu-p_m^\mu-p^\mu\right)
= \frac{E_m+E}{E_m} (2\pi\mu)^{2\ve}\frac{\d^{d-1}{\bf p}}{E}.
\eeq
Next we choose a coordinate system in the parton-parton c.m.\ frame which
has $z$ axis showing into the direction of the three-momentum ${\bf p}_Q$. 
We denote the polar and azimuthal coordinates of parton $N+1$
by $\vartheta$ and $\varphi$, so the four-vector $p^\mu$ in $d$
dimensions is
\beq
p^\mu = E
(1,\cos\vartheta,\sin\vartheta\cos\varphi,\sin\vartheta\sin\varphi,\ldots),
\quad 0<\vartheta,\varphi<\pi.
\eeq
The first four components are energy, $z$, $x$ and $y$ components of 
the three-momentum and the dots mean $d-4$ unspecified components. 
In this coordinate system
\beqn
&&(2\pi\mu)^{2\ve}\frac{\d^{d-1}{\bf p}}{E}=
E\left(\frac{2\pi\mu}{E}\right)^{2\ve}\d E\,(\sin \vartheta)^{(1-2\ve)}
\d\vartheta \,d^{1-2\ve}\varphi
\\ \nn &&\qquad\qquad\qquad\:\equiv
E\left(\frac{2\pi\mu}{E}\right)^{2\ve}\d E\,(\sin \vartheta)^{(1-2\ve)}
\d\vartheta \,(\sin \varphi)^{-2\ve}\d \varphi \,\d^{-2\ve}\Omega
\eeqn
with $\Omega$ being the solid angle in $d-4$ dimensions.
We change integration variables from $(E,\cos\vartheta)$ to
$(z,\cos\omega)$, where $z = E_m/(E_m+E)$, so
\beq
\label{Ezrelation}
E=\frac{1-z}{z}E_m
\eeq
and $\omega$ is the angle
between the three-momenta ${\bf p}_m$ and $\bf p$ in this coordinate
system,
\beq
\cos\vartheta = \frac{1-z+z\cos\omega}{\rho},
\eeq
with $\rho$ being the ratio $p_Q/(E_m+E)$ that can be expressed in
terms of $z$ and $\cos\omega$ as
\beq
\rho = \sqrt{1-2 z(1-z)(1-\cos\omega)}.
\eeq
In the collinear limit of particles $m$ and $N+1$, i.e. 
$\omega\to 0$, the definition of $z$ given here is identical to 
the one given in the previous subsection and therefore, extends this 
variable to non-collinear configurations naturally. We remark that 
this definition of $z$ is not boost invariant.

Finally, we change variable from $\omega$ to $Q^2=s_{m,N+1}$.
The necessary relation is
\beq
\label{omegaQ2relation}
\cos\omega = 1-\frac{Q^2}{2 z (1-z) E_Q^2}
\eeq
with $E_Q^2$ being dependent on $Q^2$. We obtain
\beqn
\label{measurefin}
&&\D_{N+1}^{\fin;m}(\ve) =
\Bigg[\D_N(\ve)\,
\delta^{(4)}\left(p_A^\mu+p_B^\mu-\sum_{i=1}^{N} p_i^\mu\right)
\Bigg]\st{m\to Q} 
\\ \nn && \qquad\qquad\times 
\frac{1}{2\rho^{5-4\ve}} \left(\frac{p_Q^2}{E_Q^2} \right)^2
\left(\frac{2\pi\mu}{p_Q}\right)^{2\ve}[z(1-z)]^{-\ve}\d z\,
\\ \nn && \qquad\qquad\times
\left(\frac{Q^2}{2E_Q^2}\right)^{-\ve}
\left(2-\frac{Q^2}{2z(1-z)E_Q^2}\right)^{-\ve}\d Q^2
\\ \nn && \qquad\qquad\times
(\sin\varphi)^{-2\ve}\d \varphi\,\d^{-2\ve}\Omega\,
\Theta(4 z (1-z)E_Q^2>Q^2)
\\ \nn && \qquad\qquad\times
\Theta(r^{(N+1;i)}_{\rm min} > r^{(N+1;f)}_{\rm min})
\Theta(s^{(N;f)}_{\rm min} > \s k{,N+1}^{\rm min}) \Theta(E_k > E_{N+1})
\\ \nn &&\qquad\qquad\times
\Theta(\min(r_{i,N+1}: i=1,\ldots,N, ~ i\ne m) > r_{m,N+1}),\quad
\varphi\in[0,\pi].
\eeqn
In this equation the $p_m^\mu=p_Q^\mu-p^\mu$ 
momentum conservation constraint is implicitly understood.
We shall use the four dimensional limit of this integration measure:
\beqn
\label{measurefin4}
&&\D_{N+1}^{\fin;m}(\ve=0) =
\Bigg[\D_N(\ve=0)\,
\delta^{(4)}\left(p_A^\mu+p_B^\mu-\sum_{i=1}^{N} p_i^\mu\right)
\Bigg]\st{m\to Q} 
\\ \nn && \qquad\qquad\times 
\frac{1}{2\rho^5} \left(\frac{p_Q^2}{E_Q^2} \right)^2
\d z\,\d Q^2\,\d \varphi\,\Theta(4 z (1-z)E_Q^2>Q^2)
\\ \nn && \qquad\qquad\times
\Theta(r^{(N+1;i)}_{\rm min} > r^{(N+1;f)}_{\rm min})
\Theta(s^{(N;f)}_{\rm min} > \s k{,N+1}^{\rm min}) \Theta(E_k > E_{N+1})
\\ \nn &&\qquad\qquad\times
\Theta(\min(r_{i,N+1}: i=1,\ldots,N, ~ i\ne m) > r_{m,N+1}),\quad
\varphi\in[-\pi,\pi].
\eeqn

We now turn to the discussion of the integration measure 
$\D_{N+1}^{\ini;A}$ . In order to write it in the
required form, we imagine a $2\to 2$ scattering, $A+B\to Q+(N+1)$, 
followed by the decay of particle $Q$ into $N$ particles. We write the 
integration measure $\D_{N+1}^{\ini;A}$ in the form 
\beqn
\label{measureini}
&&\D_{N+1}^{\ini;A}(\ve)=
\D_N(\ve)
\delta^d\left(p_Q^\mu-\sum_{i=1}^N p_i^\mu\right)\st{p_Q^\mu
=p_A^\mu+p_B^\mu-p^\mu}
(2\pi\mu)^{2\ve} \d^{4-2\ve} p\,2\,\delta\!\left(p^\mu p_\mu\right)
\\ \nn && \qquad\qquad\times
\Theta(r^{(N+1;f)}_{\rm min} > r^{(N+1;i)}_{\rm min})
\Theta(s^{(N;i)}_{\rm min} > s_{X,N+1}^{\rm min})
\Theta(r_{B,N+1} > r_{A,N+1}).
\eeqn
For the invariant measure of particle $N+1$ we use the the variables 
$\xi$ and {\bf W} (introduced in ref.\ \cite{EKSjets}) that are defined 
in the {\em hadron-hadron c.m. frame} such that the four-momentum of 
particle $N+1$ in light cone coordinates ($p^\mu=(p^+,p^-,{\bf p})$, 
$p^\pm = (p^0\pm p^3)/\sqrt{2})$) can be written as
\beq
\label{xiW}
p^\mu_{N+1} = \left(\xi\sqrt{\frac{s}{2}},\frac{\xi W^2}{\sqrt{2s}},\xi
{\bf W}\right).
\eeq
In these variables, the soft limit is controlled by $\xi\to 0$, while the
limit when particle $N+1$ becomes collinear to particle $A$ is controlled 
by {\bf W}$\to${\bf 0} and the invariant measure becomes
\beqn
&&(2\pi\mu)^{2\ve}
\d^{4-2\ve} p_{N+1}\,2\,\delta\!\left(p_{N+1}^2\right) =
\xi^{1-2\ve}\d\xi\,(2\pi\mu)^{2\ve}\d^{2-2\ve}{\bf W}
\\ \nn &&\qquad\qquad\qquad\qquad\qquad\qquad \:\equiv 
\xi W\left(\frac{2\pi\mu}{\xi W}\right)^{2\ve}\d \xi\,
\d W \,(\sin \phi)^{-2\ve}\d \phi \,\d^{-2\ve}\Omega,
\eeqn
where $W=|{\bf W}|$ and $\phi$ is the azimuthal angle of particle $N+1$.
In light cone coordinates the incoming particles have four-momenta
\beq
p_A^\mu = \left(x_A\sqrt{\frac{s}{2}},0,{\bf 0}\right), \quad
p_B^\mu = \left(0,x_B\sqrt{\frac{s}{2}},{\bf 0}\right),
\eeq
hence, $\xi<x_A$. The invariant mass squared of particle $Q$ has to be 
greater then zero, which constraints the upper value of $\xi W^2$,
$\xi W^2 < x_Q x_B s/x_A$ with $x_Q = x_A-\xi$. For later use we 
record the four-dimensional limit of the integration measure 
$\D_{N+1}^{\ini;A}$:
\beqn
\label{measureini4}
&&\D_{N+1}^{\ini;A}(\ve=0)=
\D_N(\ve=0)
\delta^{(4)}\left(p_Q^\mu-\sum_{i=1}^N p_i^\mu\right)\st{p_Q^\mu
=p_A^\mu+p_B^\mu-p^\mu}
\frac{1}{2}\xi\d\xi\,\d W^2\,\d \phi
\\ \nn && \qquad\qquad\times
\Theta(r^{(N+1;f)}_{\rm min} > r^{(N+1;i)}_{\rm min})
\Theta(s^{(N;i)}_{\rm min} > s_{X,N+1}^{\rm min})
\Theta(r_{B,N+1} > r_{A,N+1}).
\eeqn

The decomposition of the phase space into initial and final pieces
(see eq.\ (\ref{psdecomp})) naturally decomposes the subtraction 
terms and the finite contribution as well,
\beqn
&&\I[\soft]_m = \sum_{m_f=1}^N\I[\soft]_m^{m_f} 
+ \sum_{m_i=A,B}\I[\soft]_m^{m_i},  \\ 
&&\I[\coll]_m = \sum_{m_f=1}^N\I[\coll]_m^{m_f} 
+ \sum_{m_i=A,B}\I[\coll]_m^{m_i}, \\
\label{Ifin}
&&\I[\fin] = \sum_{m_f=1}^N\I[\fin]^{m_f} 
+ \sum_{m_i=A,B}\I[\fin]^{m_i}.
\eeqn
In the following subsections, we define these terms precisely.

\subsection{Soft subtractions}

In this subsection we define the $\I[\soft]_m^x$ integrals for the
cases $m=A,B,1,\ldots,N$ and $x=A,B,1,\ldots,N$. We start 
with the integrals involving the measure $\D_{N+1}^{\fin;m}$.

In the soft limit, the second and third $\Theta$ functions in eq.\
(\ref{identity}) take the value one and so does the factor $(E_m+E)/E_m$ in
eq.\ (\ref{pmintegral}). We find
\beqn
\label{measurefinsoft}
&&\lim_{p^\mu\to 0}\D_{N+1}^{\fin;m}(\ve) =
\Bigg[\D_N(\ve)\,(2\pi\mu)^{-2\ve}
\delta^d\left(p_A^\mu+p_B^\mu-\sum_{i=1}^{N} p_i^\mu\right)
\Bigg]\st{m\to P}
\\ \nn && \qquad\qquad\qquad\quad\times 
(2\pi\mu)^{2\ve} \d^{4-2\ve} p\,2\,\delta\!\left(p^\mu p_\mu\right)\,
\Bigg[\Theta(r^{(N+1;i)}_{\rm min} > r^{(N+1;f)}_{\rm min})
\\ \nn && \qquad\qquad\qquad\quad\times 
\Theta(\min(r_{i,N+1}: i=1,\ldots,N, ~ i\ne m) > r_{m,N+1})
\Bigg]\st{m_f\to P}.
\eeqn
We shall also use the four-dimensional limit of this measure expressed in
terms of $z$ and $Q^2$:
\beqn
\label{measurefin4soft}
&&\lim_{p^\mu\to 0}\D_{N+1}^{\fin;m}(\ve=0) =
\Bigg[\D_N(\ve=0)\,
\delta^{(4)}\left(p_A^\mu+p_B^\mu-\sum_{i=1}^{N} p_i^\mu\right)
\Bigg]\st{m\to P} 
\\ \nn && \qquad\qquad\times 
\frac{1}{2}
\d z\,\d Q^2\,\d \varphi\,\Theta(4 E_P^2 (1-z) \ge Q^2)
\Theta(r^{(N+1;i)}_{\rm min} > r^{(N+1;f)}_{\rm min})
\\ \nn && \qquad\qquad\times 
\Theta(\min(r_{i,N+1}: i=1,\ldots,N, ~ i\ne m) > r_{m,N+1}).
\eeqn
This expression is consistent with the soft limit of eq.\
(\ref{measurefin4}). In the soft limit, $z$ and $Q^2$ are related to the
energy and polar angle according to the relations
\beq
\label{softrelations}
z=1-\frac{E}{E_P},\quad Q^2=2 (1-z) E_P^2 (1-\cos\vartheta).
\eeq

Using the soft limit of $\D_{N+1}^{\fin;m}(\ve)$, we define the soft 
terms $\I[\soft]_m^{m_f}$ ($m = A,B,1,\ldots,N$) as
\newpage
\beqn
\label{Isoftmf}
&&\I[\soft]_m^{m_f} =
\\ \nn &&\sum_{a_A,a_B,a_1,\ldots,a_N}
\int \frac{\as}{(2\pi)^2} \Bigg[\D_N(\ve)\, (2\pi\mu)^{-2\ve} 
\delta^d\left(p_A^\mu+p_B^\mu-\sum_{i=1}^N p_i^\mu\right)
\Bigg]\st{m_f\to P}
\\ \nn &&\qquad\qquad\qquad \times 
(2\pi\mu)^{2\ve} \d^{4-2\ve} p\,2\,\delta\!\left(p^\mu p_\mu\right)\,
\Bigg[\Theta(r^{(N+1;i)}_{\rm min} > r^{(N+1;f)}_{\rm min})
\\ \nn &&\qquad\qquad\qquad \times 
\Theta(\min(r_{i,N+1}: i=1,\ldots,N, ~ i\ne m_f) > r_{m_f,N+1})
\\ \nn &&\qquad\qquad\qquad \times 
\sum_{{n=A,B,1,\ldots,N} \choose n\ne m}
\Psi^{(2N+2)}_{S;mn}(\vec{a},\vec{p},p_{N+1}^\mu)
\Theta(\alpha\s mn > \s m{,N+1}+\s n{,N+1})
\Bigg]\st{m_f\to P}
\\ \nn &&\qquad\qquad\qquad \times 
L(a_A,a_B,x_A,x_B)\S_{N+1}(p_1^\mu,\ldots,p^\mu)
\st{{m_f\to P}\choose {p^\mu=0}},\qquad 0<\alpha<\alphamax\le 1/2.
\eeqn
The indices $m_f$ and $P$ label the same momentum, therefore, the 
$m_f\to P$ substitution is formal in this equation and we shall omit it
in the rest of the paper.
In eq.\ (\ref{Isoftmf}), we introduced an upper bound --- $\alpha$ times
the energy of parton $m$ --- for the energy of parton $N+1$ in the c.m.\ 
frame of partons $m$ and $n$, expressed in terms of invariants. Physical 
quantities will not depend on this bound. The $z=1-E/E_P>0$ relation
introduces an upper bound on $E$ which is not present in
eq.\ (\ref{measurefinsoft}). Thus the use of the soft approximation is
justified only if 
\beqn
&&\Theta(\alpha\s mn > \s m{,N+1}+\s n{,N+1})\Theta(E_P>E)
S_{N+1}(\dc)\st{p_{N+1}^\mu=0}=
\\ \nn &&\qquad\qquad
\Theta(\alpha\s mn > \s m{,N+1}+\s n{,N+1})
S_{N+1}(\dc)\st{p_{N+1}^\mu=0}
\eeqn
for any $m$ and $n$.
When calculating an $N$-jet quantity at a fixed value of $\dc$, there
always exist a finite $\alphamax$ such that this equation is fulfilled
if $\alpha<\alphamax$. The value of $\alphamax$ is easily obtained by
integrating the measurement function only.

We now turn to the definition of the soft subtraction terms involving 
the measure $\D_{N+1}^{\ini;A}$. The soft limit, $p^\mu\to 0$ is
equivalent to taking $\xi\to 0$, in which case eq.\ (\ref{measureini}) 
takes the form
\beqn
\label{measureinisoft}
&&\lim_{\xi\to 0}\D_{N+1}^{\ini;A}(\ve) =
\D_N(\ve)(2\pi\mu)^{-2\ve}
\delta^d\left(p_A^\mu+p_B^\mu-\sum_{i=1}^N p_i^\mu\right)
\\ \nn && \qquad\qquad\qquad\times
\xi W\left(\frac{2\pi\mu}{\xi W}\right)^{2\ve}\d \xi\,
\d W \,(\sin \phi)^{-2\ve}\d \phi \,\d^{-2\ve}\Omega
\\ \nn && \qquad\qquad\qquad\times
\Theta(r^{(N+1;f)}_{\rm min} > r^{(N+1;i)}_{\rm min})
\Theta(r^{B,N+1} > r^{A,N+1}).
\eeqn
The second line contains the invariant measure of particle $N+1$,
so we define the soft terms $\I[\soft]_m^{A}$ ($m = A,B,1,\ldots,N$) as
\newpage
\beqn
\label{Isoftmi}
&&\I[\soft]_m^{A} =
\\ \nn &&\sum_{a_A,a_B,a_1,\ldots,a_N}
\int \frac{\as}{(2\pi)^2} \D_N(\ve)\,
(2\pi\mu)^{-2\ve} 
\delta^d\left(p_A^\mu+p_B^\mu-\sum_{i=1}^N p_i^\mu\right)
\\ \nn &&\qquad\qquad\qquad \times 
(2\pi\mu)^{2\ve} \d^{4-2\ve} p\,2\,\delta\!\left(p^\mu p_\mu\right)\,
\\ \nn &&\qquad\qquad\qquad \times 
\Theta(r^{(N+1;f)}_{\rm min} > r^{(N+1;i)}_{\rm min})
\Theta(r^{B,N+1} > r^{A,N+1})
\\ \nn &&\qquad\qquad\qquad \times 
\sum_{{n=A,B,1,\ldots,N} \choose n\ne m}
\Psi^{(2N+2)}_{S;mn}(\vec{a},\vec{p},p_{N+1}^\mu)
\Theta(\alpha\s mn > \s m{,N+1}+\s n{,N+1})
\\ \nn &&\qquad\qquad\qquad \times 
L(a_A,a_B,x_A,x_B)\S_{N+1}(p_1^\mu,\ldots,p^\mu)\st{p^\mu\to 0},\qquad
0<\alpha<\alphamax .
\eeqn
Here, it is useful to choose $\alphamax$ such that
\beqn
&&\Theta(x_A>\xi)\Theta((x_A-\xi)x_B s>x_A\xi W^2)
\\ \nn &&\qquad\qquad\times
\Theta(\alpha\s mn > \s m{,N+1}+\s n{,N+1})
S_{N+1}(\dc)\st{p_{N+1}^\mu=0}
\\ \nn &&\qquad\qquad
=\Theta(\alpha\s mn > \s m{,N+1}+\s n{,N+1})
S_{N+1}(\dc)\st{p_{N+1}^\mu=0}
\eeqn
for any $m$ and $n$. In this case the generation of the soft phase space
in (\ref{IfinA}) becomes simpler.

The definition of the soft term $\I[\soft]_m^B$ is analogous. One
simply interchanges labels $A$ and $B$ in the second $\Theta$ function 
of the third line of eq.\ (\ref{Isoftmi}).
 We remark that the sum of the $N+2$ soft terms 
$\I[\soft]_m^{m_f}$ and $\I[\soft]_m^{m_i}$ is independent of the phase
space decomposition:
\beqn
\label{Isoftm}
&&\I[\soft]_m =
\\ \nn &&\sum_{a_A,a_B,a_1,\ldots,a_N}
\int \frac{\as}{(2\pi)^2} \D_N(\ve)\,
(2\pi\mu)^{-2\ve} 
\delta^d\left(p_A^\mu+p_B^\mu-\sum_{i=1}^N p_i^\mu\right)
\\ \nn &&\qquad\qquad\qquad \times 
(2\pi\mu)^{2\ve} \d^{4-2\ve} p\,2\,\delta\!\left(p^\mu p_\mu\right)
\\ \nn &&\qquad\qquad\qquad \times 
\sum_{{n=A,B,1,\ldots,N} \choose n\ne m}
\Psi^{(2N+2)}_{S;mn}(\vec{a},\vec{p},p_{N+1}^\mu)
\Theta(\alpha\s mn > \s m{,N+1}+\s n{,N+1})
\\ \nn &&\qquad\qquad\qquad \times 
L(a_A,a_B,x_A,x_B)\S_{N+1}(p_1^\mu,\ldots,p^\mu)\st{p^\mu\to 0}.
\eeqn
This relation will be the starting point for the evaluation of the
integrals over the invariant measure of particle $N+1$ in the soft
subtraction terms $\I[\soft]_m$.

We close the definition of the soft subtraction terms with spelling
out the four-dimensional limit of the measure (\ref{measureinisoft}),
\beqn
\label{measureini4soft}
&&\lim_{\xi\to 0}\D_{N+1}^{\ini;A}(\ve=0) =
\D_N(\ve=0)
\delta^{(4)}\left(p_A^\mu+p_B^\mu-\sum_{i=1}^N p_i^\mu\right)
\frac{1}{2}\xi \d \xi\, \d W^2 \,\d \phi
\\ \nn && \qquad\qquad\qquad\qquad\times
\Theta(r^{(N+1;f)}_{\rm min} > r^{(N+1;i)}_{\rm min})
\Theta(r^{B,N+1} > r^{A,N+1})
\eeqn
This form is consistent with the soft limit of eq.\ (\ref{measureini4}).
It will be used for giving an explicit expression for the finite 
integral $\I[\fin]$ in four dimensions.

\subsection{Collinear subtractions}

In this subsection we define the $\I[\coll]_m^x$ integrals for the
cases $m=A,B,1,\ldots,N$ and $x=A,B,1,\ldots,N$. We start 
with the integrals involving the measure $\D_{N+1}^{\ini;X}$ and
we discuss in detail the case $X=A$. The treatment of case $X=B$ is
analogous.

In the collinear limit $p^\mu_P=z p^\mu_A$, $p^\mu=(1-z) p^\mu_A$, the 
first $\Theta$ function in the integration measure (\ref{measurefin}) 
is zero, therefore, the terms $\I[\coll]_A^x$ for $x=1,\ldots,N$ are 
defined to be zero. Also in this limit, the $\Theta$ function 
$\Theta(r_{A,N+1}>r_{B,N+1})$ appearing in the measure
$\D_{N+1}^{\ini;B}$ becomes zero, therefore, the term $\I[\coll]_A^B$ 
is defined to be zero. In the same limit the measure 
$\D_{N+1}^{\ini;A}(\ve)$ becomes
\beq
\label{measureinicoll}
\lim_{W\to 0}\D_{N+1}^{\ini;A}(\ve)=
\D_N(\ve)(2\pi\mu)^{-2\ve}
\delta^d\left(z p_A^\mu + p_B^\mu-\sum_{i=1}^N p_i^\mu\right)
\xi \left(\frac{2\pi\mu}{\xi}\right)^{2\ve}\d \xi\,\d^{2-2\ve}{\bf W}.
\eeq
where the momentum fraction $z$ can be expressed in terms of actual
integration variables $x_A$ and $\xi$ using the momentum conservation 
relation for the ``+'' component of the momenta,
\beq
\label{zini}
z = \frac{x_A - \xi}{x_A}.
\eeq
The collinear pole $2/s_{A,N+1}$ equals $2/(x_A\xi W^2)$, therefore,
the collinear term $\I[\coll]_A^A$ is defined as
\beqn
\label{collAA}
&&\I[\coll]_A^A =
\\ \nn &&\sum_{a_A,a_B,a_1,\ldots,a_N}
\int \frac{\as}{2\pi^2} \D_N(\ve) L(a_A,a_B,x_A,x_B)
\left(\frac{2\pi\mu}{\xi}\right)^{2\ve}\d \xi\,
\frac{\d^{2-2\ve}{\bf W}}{x_A W^2}
\\ \nn &&\qquad\qquad\quad \times 
\Bigg[(2\pi\mu)^{-2\ve}
\delta^d\left(z p_A^\mu + p_B^\mu-\sum_{i=1}^N p_i^\mu\right)
\Theta(z s\delta/x_A>W^2)\Theta(x_A>\xi)
\\ \nn &&\qquad\qquad\qquad\quad \times 
\sum_{a_{N+1}}
(-1)^{1+f(a_A)+f(a_P)}\Psi_{C;\bar{A},N+1}^{(2N+2)}(1/z,P;A,B,1,\ldots,N+1) 
\\ \nn &&\qquad\qquad\qquad\quad \times 
\S_{N+1}(p_1^\mu,\ldots,p^\mu)\st{p^\mu=(1-z)p_A^\mu}
\\ \nn &&\qquad\qquad\qquad
- (2\pi\mu)^{-2\ve}
\delta^d\left(p_A^\mu + p_B^\mu-\sum_{i=1}^N p_i^\mu\right)
\Theta(s\delta/x_A>W^2)\Theta(z>1-\alpha)
\\ \nn &&\qquad\qquad\qquad\quad \times 
\frac{2C(a_A)}{(1-z)} \Psi^{(2N)}(\vec{a},\vec{p})
\S_{N+1}(p_1^\mu,\ldots,p^\mu)\st{p^\mu=0}\Bigg],\quad
0<\delta<\deltamax<1,
\eeqn
where we introduced a convenient upper bound for the $W$ integral.
We have subtracted a term in the soft-collinear limit ($z\to 1$) in order
to keep $\I[\coll]_A^A$ from having a soft divergence when $\ve\to 0$.
According to eq.\ (\ref{zini}), in this subtracted term the $z>1-\alpha$
constraint is equivalent to $\alpha x_A > \xi$ and $1/(1-z) = x_A/\xi$.
It is useful to choose $\deltamax$ such that 
\beq
\Theta((x_A-\xi)x_B s > z \xi s \deltamax)
S_{N+1}(\dc)\st{p_{N+1}^\mu=(1-z)p_A^\mu}=
S_{N+1}(\dc)\st{p_{N+1}^\mu=(1-z)p_A^\mu},
\eeq
in which case the generation of the collinear phase space
in (\ref{IfinA}) becomes simpler.

We shall also use the four-dimensional limit of measure
(\ref{measureinicoll})
\beqn
\label{measureini4coll}
&&\lim_{W\to 0}\D_{N+1}^{\ini;A}(\ve=0)=
\D_N(\ve=0)
\delta^{(4)}\left(z p_A^\mu + p_B^\mu-\sum_{i=1}^N p_i^\mu\right)
\frac{1}{2}\xi d \xi\, \d W^2 \,\d \phi.
\eeqn
This form is consistent with the collinear limit of eq.\
(\ref{measureini4}). It will be used for giving an explicit expression
for the finite integral $\I[\fin]$ in four dimensions.

Next we consider the collinear subtraction terms involving the measure
$\D_{N+1}^{\fin;m}$. The collinear limit of particles $m$ and $N+1$
implies taking $\omega_{m,N+1}\equiv\omega\to 0$, $Q^2\to 0$. In this
limit, in the integration 
measure (\ref{measurefin}) the first two $\Theta$ functions become one,
the third one becomes $\Theta(z > 1/2)$ and the last one becomes 
$\delta_{mm_f}$. We find
\beqn
\label{measurefincoll}
&&\lim_{\omega,Q^2\to 0} \D_{N+1}^{\fin;m_f}(\ve) =
\Bigg[\D_N(\ve)\,(2\pi\mu)^{-2\ve}
\delta^d\left(p_A^\mu+p_B^\mu-\sum_{i=1}^{N} p_i^\mu\right)
\Bigg]\st{m_f\to P} \delta_{mm_f}
\\ \nn && \qquad\qquad\times 
\frac{1}{2}
\left(\frac{2\pi\mu}{E_P}\right)^{2\ve}[z(1-z)]^{-\ve}\d z\,
\left(\frac{Q^2}{E_P^2}\right)^{-\ve}\d Q^2\,
(\sin\varphi)^{-2\ve}\d \varphi\,\d^{-2\ve}\Omega
\\ \nn && \qquad\qquad\times 
\Theta(4 z (1-z)E_P^2>Q^2)\,\Theta(z > 1/2).
\eeqn
The collinear pole $2/s_{m_f,N+1}=2/Q^2$, so we define the collinear terms 
$\I[\coll]_m^{m_f}$ as
\beqn
\label{Icollmf}
&&\I[\coll]_m^{m_f} =
\\ \nn &&\sum_{a_A,a_B,a_1,\ldots,a_N}
\int \frac{\as}{(2\pi)^2}\Bigg[\D_N(\ve) (2\pi\mu)^{-2\ve} 
\delta^d\left(p_A^\mu+p_B^\mu-\sum_{i=1}^N p_i^\mu\right)
\Bigg]\st{m_f\to P} \delta_{mm_f}
\\ \nn &&\qquad\qquad\qquad \times 
L(a_A,a_B,x_A,x_B)\left(\frac{2\pi\mu}{E_P}\right)^{2\ve}(1-z)^{-\ve}\d z
\left(\frac{Q^2}{E_P^2}\right)^{-\ve}\frac{\d Q^2}{Q^2}
\d^{1-2\ve}\varphi
\\ \nn &&\qquad\qquad\qquad \times 
\Bigg[z^{-\ve}\,\Theta(4 z (1-z)E_P^2\delta>Q^2)\,
\\ \nn &&\qquad\qquad\qquad\qquad \times 
\sum_{a_{N+1}}\Psi_{C;m_f,N+1}^{(2N+2)}(z,P;A,B,1,\ldots,N+1)
\\ \nn &&\qquad\qquad\qquad\qquad \times 
\Theta(z > 1/2) \S_{N+1}(p_1^\mu,\ldots,p^\mu)
\st{{p_m^\mu=z p_P^\mu}\choose {p^\mu=(1-z) p_P^\mu}}
\\ \nn &&\qquad\qquad\qquad \quad
-\Theta(4 (1-z)E_P^2\delta>Q^2)\,
\frac{2C(a_m)}{1-z}\Psi^{(2N)}(\vec{p})\st{m_f\to P}
\\ \nn &&\qquad\qquad\qquad \qquad
\Theta(z>1-\alpha)\,
\S_{N+1}(p_1^\mu,\ldots,p^\mu)
\st{{p_m^\mu=p_P^\mu}\choose {p^\mu=0}}
\Bigg],\quad 0<\delta<\deltamax.
\eeqn
Here we have subtracted the integrand at $z=1$ in order to keep 
$\I[\coll]_m^{m_f}$ from having a soft divergence when $\ve\to 0$.
After making the indicated substitutions, the right hand side does not 
contain the indices $m$ and $m_f$, but the index $P$. 
In equation (\ref{Icollmf}), for the case $\delta=1$ the upper limit
on the $Q^2$ integration derives from the $\cos\omega \ge -1$ constraint
with the use of the relations
\beq
Q^2 = 2z(1-z)E_P^2(1-\cos\omega),\qquad
Q^2 = 2(1-z)E_P^2(1-\cos\omega)
\eeq
in the collinear and soft-collinear limits respectively. 
We remind the reader that particle $P$ is the massless limit of particle
$Q$. In the collinear limit $p_m^\mu=z p_P^\mu$, $p^\mu\to (1-z)p_P^\mu$ 
the first $\Theta$ function in the measure (\ref{measureini}) becomes zero,
therefore, the collinear subtraction terms $\I[\coll]_m^x$ for $x=A,B$ 
are defined to be zero, therefore,
\beq
\I[\coll]_m = \sum_{m_f=1}^N \I[\coll]_m^{m_f}.
\eeq

Finally we spell out the four-dimensional limit of the measure
(\ref{measurefincoll}), which is consistent with the collinear
limit of the measure (\ref{measurefin4}):
\beqn
\label{measurefin4coll}
&&\lim_{\omega\to 0}\D_{N+1}^{\fin;m}(\ve=0) =
\Bigg[\D_N(\ve=0)\,
\delta^{(4)}\left(p_A^\mu+p_B^\mu-\sum_{i=1}^{N} p_i^\mu\right)
\Bigg]\st{m\to P} 
\\ \nn && \qquad\qquad\times 
\frac{1}{2}
\d z\,\d Q^2\,\d \varphi\,\Theta(4 z(1-z) E_P^2 \ge Q^2)\Theta(z>1/2).
\eeqn

\subsection{The finite contribution}

The precise definition of the soft and collinear terms fixes the finite
term by eqs.\ (\ref{2toN+1}) and (\ref{2toN+1decomp}). For the sake of
completeness, in this subsection we write it in terms of the actual 
integration variables. The integrand contains at most integrable
square-root singularities over the integration domain, therefore, 
it suffices to write the integral in $d=4$ dimensions.
We give the integrals $\I[\fin]^{m_i}$ explicitly for the case $m_i=A$
only.
\beqn
\label{IfinA}
&&\I[\fin]^A\st{\ve=0} =
\sum_{a_A,a_B,a_1,\ldots,a_N}
\int \frac{\as}{(2\pi)^2}
\D_N(0)\frac{1}{2}\xi \d \xi\,\d W^2\,\d \phi\,L(a_A,a_B,x_A,x_B)
\\ \nn &&\qquad\qquad\times
\Bigg\{
\delta^{(4)}\left(p_A^\mu + p_B^\mu - \sum_1^N p_i^\mu\right)
\Theta(r^{(N+1;f)}_{\rm min} > r^{(N+1;i)}_{\rm min})
\\ \nn &&\qquad\qquad\qquad\times
\Theta(x_A > \xi)\Theta((x_A-\xi)x_B s> x_A\xi W^2)
\\ \nn &&\qquad\qquad\qquad\times
\Theta(s^{(N;i)}_{\rm min} > \s X{,N+1}^{\rm min}) 
\Theta(r_{B,N+1} > r_{A,N+1})
\\ \nn &&\qquad\qquad\qquad\times
\sum_{a_{N+1}}\Psi^{(2N+2)}(A,B,1,\ldots,N+1)
\S_{N+1}(p_1^\mu,\ldots,p_{N+1}^\mu)
\\ \nn &&\qquad\qquad\quad 
- \delta^{(4)}\left(p_A^\mu + p_B^\mu - \sum_1^N p_i^\mu\right)
\Theta(r^{(N+1;f)}_{\rm min} > r^{(N+1;i)}_{\rm min})
\Theta(r_{B,N+1} > r_{A,N+1})
\\ \nn &&\qquad\qquad\qquad \times
\sum_{{m,n=A,B,1,\ldots,N} \choose m<n}
\frac{2 \s mn}{\s m{,N+1}\s n{,N+1}} \Psi^{(2N;c)}_{mn}(\vec{a},\vec{p})
\Theta(\alpha \s mn > \s m{,N+1}+\s n{,N+1})
\\ \nn &&\qquad\qquad\qquad \times
\S_{N+1}(p_1^\mu,\ldots,p^\mu)\st{p^\mu=0}
\\ \nn &&\qquad\qquad\quad
- \delta^{(4)}\left(z p_A^\mu+p_B^\mu-\sum_1^N p_i^\mu\right)
\Theta(z s\delta/x_A>W^2) \Theta(x_A>\xi)
\\ \nn &&\qquad\qquad\qquad \times
\frac{2}{x_A\xi W^2} \sum_{a_{N+1}} (-1)^{1+f(a_A)+f(a_P)}
\\ \nn &&\qquad\qquad\qquad \times
\Psi_{C;A,N+1}^{(2N+2)}(1/z,P;A,B,1,\ldots,N+1)
\S_{N+1}(p_1^\mu,\ldots,p^\mu)\st{p^\mu=(1-z) p_A^\mu}
\\ \nn &&\qquad\qquad\quad
+ \delta^{(4)}\left(p_A^\mu + p_B^\mu - \sum_1^N p_i^\mu\right)
\Theta(s\delta /x_A>W^2) \Theta(\alpha x_A > \xi)
\\ \nn &&\qquad\qquad\qquad \times
2C(a_A)\frac{2}{\xi^2 W^2} \Psi^{(2N)}(\vec{a},\vec{p})
\S_{N+1}(p_1^\mu,\ldots,p^\mu)\st{p^\mu=0}
\Bigg\}
\eeqn
with $\phi\in [-\pi,\pi]$ and $z$ given in eq.\ (\ref{zini}).
The integral $\I[\fin]^B$ is analogous with the labels $A$ and $B$ and the
momentum components $p_{N+1}^+$ and $p_{N+1}^-$ interchanged.
The $\Theta$ functions assure that the integrand can become
singular only when $\xi W^2\to 0$. For the reconstruction of 
the four-momentum $p_{N+1}^\mu$ see eq.\ (\ref{xiW}).

The explicit form of the integral $\I[\fin]^{m_f}$ is
\beqn
\label{Ifinf}
&&\I[\fin]^{m_f}\st{\ve=0} =
\sum_{a_A,a_B,a_1,\ldots,a_N}
\int \frac{\as}{(2\pi)^2}
\frac{1}{2}\d z\,\d Q^2\,\d \varphi\,L(a_A,a_B,x_A,x_B)
\\ \nn &&\qquad\qquad\times
\Bigg\{
\frac{1}{\rho^5}\left(\frac{p_Q^2}{E_Q^2}\right)^2\Bigg[\D_N(0)
\delta^{(4)}\left(p_A^\mu + p_B^\mu - \sum_1^N p_i^\mu\right)\Bigg]\st{m_f\to Q}
\\ \nn &&\qquad\qquad\qquad\times
\Theta(\shat > Q^2) \Theta(4 z (1-z) E_Q^2 > Q^2)
\\ \nn &&\qquad\qquad\qquad\times
\Theta(r^{(N+1;i)}_{\rm min} > r^{(N+1;f)}_{\rm min})
\Theta(s^{(N;f)}_{\rm min} > \s k{,N+1}^{\rm min}) \Theta(E_k > E_{N+1})
\\ \nn &&\qquad\qquad\qquad\times
\Theta(\min(r_{i,N+1}: i=1,\ldots,N, ~ i\ne m_f) > r_{m_f,N+1})
\\ \nn &&\qquad\qquad\qquad\times
\sum_{a_{N+1}}\Psi^{(2N+2)}(A,B,1,\ldots,N+1)
\S_{N+1}(p_1^\mu,\ldots,p_{N+1}^\mu)
\\ \nn &&\qquad\qquad\quad 
- \Bigg[\D_N(0)\delta^{(4)}\left(p_A^\mu+p_B^\mu-\sum_1^N p_i^\mu\right)
\Bigg]\st{m_f\to P}
\\ \nn &&\qquad\qquad\qquad\times
\Bigg[\Theta(4(1-z)E_P^2\ge Q^2)
\Theta(r^{(N+1;i)}_{\rm min} > r^{(N+1;f)}_{\rm min})
\\ \nn &&\qquad\qquad\qquad\quad \times
\Theta(\min(r_{i,N+1}: i=1,\ldots,N, ~ i\ne m_f) > r_{m_f,N+1})
\\ \nn &&\qquad\qquad\qquad\quad \times
\sum_{{m,n=A,B,1,\ldots,N} \choose m<n}
\frac{2 \s mn}{\s m{,N+1}\s n{,N+1}} \Psi^{(2N;c)}_{mn}(\vec{a},\vec{p})
\\ \nn &&\qquad\qquad\qquad\quad \times
\Theta(\alpha \s mn > \s m{,N+1}+\s n{,N+1})
\S_{N+1}(p_1^\mu,\ldots,p^\mu)\st{p^\mu=0}
\\ \nn &&\qquad\qquad\qquad\quad
+ \Theta(4z(1-z)E_P^2\delta\ge Q^2) \Theta(z>1/2)
\\ \nn &&\qquad\qquad\qquad\qquad \times
\frac{2}{Q^2}\sum_{a_{N+1}}\Psi_{C;m_f,N+1}^{(2N+2)}(z,P;A,B,1,\ldots,N+1)
\\ \nn &&\qquad\qquad\qquad\qquad \times
\S_{N+1}(p_1^\mu,\ldots,p^\mu)
\st{{p_{m_f}^\mu=z p_P^\mu}\choose {p^\mu=(1-z) p_P^\mu}}
\\ \nn &&\qquad\qquad\qquad\quad
- \Theta(4(1-z)E_P^2\delta\ge Q^2) \Theta(z>1-\alpha)
\\ \nn &&\qquad\qquad\qquad\qquad \times
2C(a_{m_f})\frac{2}{(1-z)Q^2} \Psi^{(2N)}(\vec{a},\vec{p})
\S_{N+1}(p_1^\mu,\ldots,p^\mu)\st{p^\mu=0}
\Bigg]\Bigg\}
\eeqn
with $z\in [0,1]$ unless otherwise indicated and $\varphi\in [-\pi,\pi]$.
In these equations we used the relation
\beqn
&&\sum_{{m,n=A,B,1,\ldots,N} \choose n\ne m}
\frac{2 \s mn}{\s m{,N+1}(\s m{,N+1} + \s n{,N+1})}
\Psi^{(2N;c)}_{mn}(\vec{a},\vec{p})
\Theta(\alpha \s mn > \s m{,N+1}+\s n{,N+1})
\\ \nn &&\qquad\qquad\qquad
=\sum_{{m,n=A,B,1,\ldots,N} \choose m<n}
\frac{2 \s mn}{\s m{,N+1}\s n{,N+1}}
\Psi^{(2N;c)}_{mn}(\vec{a},\vec{p})
\Theta(\alpha \s mn > \s m{,N+1}+\s n{,N+1})
\eeqn
in order to simplify the sum of the soft subtraction terms.
The $\D_N(0)$ factor with the accompanying momentum conservation
$\delta$ function contains the $N$-body phase space, the $x_A$, $x_B$
integration (and other trivial factors). This $N$-body phase space can be
generated according to the reader's preference. For instance, one can use
the well-known phase space generating routine RAMBO \cite{rambo}. The
reconstruction of the momentum of particle $(N+1)$ from $z$, $Q^2$ and 
$\varphi$ was given in previous subsections (see formulas 
(\ref{Ezrelation}), (\ref{omegaQ2relation}) and (\ref{softrelations})).

We have given the precise definition of all terms in eq.\
(\ref{2toN+1decomp}). The integrals $\I[\soft]$ and $\I[\coll]$
contain poles in the Laurent expansion in $\ve$ around
zero. According to the factorization and Kinoshita-Lee-Nauenberg theorems,
these poles cancel against similar poles in the $\I[2\to N]$ contribution.
In order to see the cancelation of infrared divergences explicitly,
we have to analyze the integrals over the invariant measure of gluon 
$N+1$ in the soft and collinear contributions, which is the subject 
of the next two sections.

\section{Soft integrals}
\setcounter{equation}{0}

In this section, we evaluate the integrals in $\I[\soft]_m$
($m=A,B,1,\ldots,N$) over the invariant measure of particle 
$N+1$.

At the soft point, $p^\mu=0$ the measurement function
$\S_{N+1}(p_1^\mu,\ldots,p^\mu)=\S_N(p_1^\mu,\ldots,p_N^\mu)$, 
which is the manifestation of the requirement of infrared safe 
measurement. The only dependence in eq.\ (\ref{Isoftm}) on the variables 
of gluon $N+1$ is in the eikonal factor, therefore, we can write 
$\I[\soft]_m$ as
\beqn
\label{Isoftmnew}
&&\I[\soft]_m =
\sum_{a_A,a_B,a_1,\ldots,a_N}
\int \frac{\as}{(2\pi)^2} \D_N(\ve)\,
(2\pi\mu)^{-2\ve} 
\delta^d\left(p_A^\mu+p_B^\mu-\sum_{i=1}^N p_i^\mu\right)
\\ \nn &&\qquad\qquad\qquad \times 
L(a_A,a_B,x_A,x_B) \S_N(p_1^\mu,\ldots,p_N^\mu)
\sum_{{n=A,B,1,\ldots,N} \choose {n\ne m}} \J_{mn}(\vec{p})
\Psi^{(2N;c)}_{mn}(\vec{a},\vec{p}),
\eeqn
where
\beqn
&&\J_{mn}(\vec{p}) = \int
(2\pi\mu)^{2\ve}\d^{4-2\ve} p_{N+1}\,2\,\delta\!\left(p_{N+1}^2\right) 
\\ \nn &&\qquad\qquad\quad \times 
\frac{2\s mn}{\s m{,N+1}(\s m{,N+1}+\s n{,N+1})}
\Theta(\alpha\s mn > \s m{,N+1}+\s n{,N+1}).
\eeqn
This integral is evaluated in the appendix. The result can be obtained
exactly, however, for our purposes the Laurent expansion in the form,
\beq
\J_{mn}(\vec{p}) = 2\pi \cG \left(\frac{\mu^2}{\qes}\right)^\ve
\left[\frac{1}{2\ve^2}-\frac{1}{\ve}\ln \alpha 
- \frac{1}{2\ve}\ln\frac{\s mn}{\qes} + \tilde{\J}_{mn}(\vec{p}) 
+ \O(\ve)\right],
\eeq
is better suited.
The function $\tilde{\J}_{mn}$ is independent of $\ve$ and is very simple:
\beq
\label{Itilde}
\tilde{\J}_{mn}=\frac{1}{4}\ln^2\left(\alpha^2\frac{\s mn}{\qes}\right)
-\frac{\pi^2}{12}.
\eeq
Substituting this result for the $\J_{mn}$ soft integral
into eq.\ (\ref{Isoftmnew}) and using the soft-collinear identity,
formula (\ref{softcollidentity}), we see that $\I[\soft]_m$ assumes very
similar form to the $\I[2\to N]$ integral, eq.\ (\ref{I2toN}):
\beqn
&&\I[\soft]_m =
\sum_{a_A,a_B,a_1,\ldots,a_N}
\int \D_N(\ve)\S_N(p_1^\mu,\ldots,p_N^\mu) (2\pi\mu)^{-2\ve} 
\delta^d\left(p_A^\mu+p_B^\mu-\sum_{i=1}^N p_i^\mu\right)
\\ \nn &&\qquad\qquad\qquad\qquad\qquad\qquad \times 
L(a_A,a_B,x_A,x_B) \frac{\as}{2\pi}\,\cG 
\left(\frac{\mu^2}{\qes}\right)^\ve \Psi^{\soft}_m(\vec{a},\vec{p}),
\eeqn
where
\beqn
&&\Psi^{\soft}_m(\vec{a},\vec{p}) =\Psi^{(2N)}(\vec{a},\vec{p})
\left[\frac{1}{\ve^2}C(a_m)-\frac{1}{\ve}2C(a_m)\ln \alpha\right]
\\ \nn &&\qquad\qquad\quad 
+\sum_{{n=A,B,1,\ldots,N} \choose {n\ne m}} 
\Psi^{(2N;c)}_{mn}(\vec{a},\vec{p})
\left[-\frac{1}{2\ve} \ln\left(\frac{\s mn}{\qes}\right)
+ \tilde{\J}_{mn}(\vec{p})\right].
\eeqn

\section{Collinear integrals}
\setcounter{equation}{0}

In this section, we evaluate the integrals in $\I[\coll]_m^m$
($m=A,B,1,\ldots,N$) over the invariant measure of parton $N+1$.
Before going into the details, we have to make a remark. The collinear
subtraction terms were defined using the four-dimensional expressions for
the collinear limit of the squared matrix element. That was sufficient
for the evaluation of the $\I[\fin]$ integral. Strictly speaking however,
the subtraction scheme applied in this paper is defined in $d$
dimensions. It was shown in ref.\ \cite{KST2to2} that with making use of
process independent transition terms, one can use four-dimensional
expressions for the helicity independent part for the collinear limit of
the squared matrix element except for the \AP splitting functions,
$\tilde{P}_{a/b}$ that have to be calculated in $d$ dimensions.
As for the helicity dependent part, the
analysis of its general structure in $d$ dimensions shows it has
vanishing azimuthal integral in $d$ dimensions \cite{KSjets,FKSjets}.
Therefore, we drop the helicity dependent part of $\Psi_{C;mn}^{(2N+2)}$
in the following considerations. This causes some inconsistency in our
notation, but the physical cross section remains unchanged.

We start with the evaluation of the integrals in $\I[\coll]_m^m$
($m=A,B$) over the invariant measure of parton $N+1$.
At the collinear point {\bf W}$\to 0$ with $\xi$ fixed, the measurement 
function $\S_{N+1}(p_1^\mu,\ldots p^\mu) =\S_N(p_1^\mu,\ldots,p_N^\mu)$. 
We change integration variables in the first term
of eq.\ (\ref{collAA}) form $x_A$ (which is hidden in $\D_N$ and in the
luminosity factor) and $\xi$ to $x_P=x_A-\xi$ and $z$ with $z$ defined 
in eq.\ (\ref{zini}). The Jacobian for this transformation is $x_A/z$.
The lower limit for the $z$ integral is defined by the $x_A\le 1$
relation, hence $z\ge x_P$, while the upper limit is obviously one.
The limits on $x_P$ are zero and one, just as was on $x_A$. After this
change of variables we can rename the index $P$ to $A$ (and simultaneously
the flavor index $a_A$ to $b$). In the second
term of eq.\ (\ref{collAA}), we change variable from $\xi$ to $z$.
Keeping the helicity independent term, we can now write $\I[\coll]_A^A$ as
\beqn
&&\I[\coll]_A^A=
\\ \nn &&\sum_{a_A,a_B,a_1,\ldots,a_N}
\int \frac{\as}{2\pi^2}\D_N(\ve)
(2\pi\mu)^{-2\ve}\delta^d\left(p_A^\mu+p_B^\mu-\sum_{i=1}^N p_i^\mu\right)
\Psi^{(2N)}(\vec{p}) \S_N(p_1^\mu,\ldots,p_N^\mu) 
\\ \nn && \qquad\qquad\qquad \times
\int\d z\Bigg[\left(\frac{1-z}{z}x_A\right)^{-2\ve}\frac{1}{z^2}\Theta(z>x_A)
\\ \nn && \qquad\qquad\qquad \qquad \qquad \times
\sum_b \frac{\omega(b)}{\omega(a_A)}
\P_{a_A/b}(z,\ve) L\left(b,a_B,\frac{x_A}{z},x_B\right)
\\ \nn && \qquad\qquad\qquad \qquad \qquad \times
(2\pi\mu)^{2\ve} \int \frac{\d^{2-2\ve}{\bf W}}{W^2} 
\Theta\left(s\delta z^2/x_A>W^2\right)
\\ \nn && \qquad\qquad \qquad\qquad\quad
- \left((1-z)x_A\right)^{-2\ve}\Theta(z>1-\alpha)
L(a_A,a_B,x_A,x_B)\frac{2C(a_A)}{1-z}
\\ \nn && \qquad\qquad \qquad \qquad\qquad \times
(2\pi\mu)^{2\ve} \int \frac{\d^{2-2\ve}{\bf W}}{W^2} 
\Theta\left(s\delta /x_A>W^2\right)\Bigg].
\eeqn
Evaluation of the integral over {\bf W} (see eq.\ (\ref{Wint})) results in 
\beqn
&&\I[\coll]_A^A=
\\ \nn &&-\sum_{a_A,a_B,a_1,\ldots,a_N}
\int \D_N(\ve) \S_N(p_1^\mu,\ldots,p_N^\mu) 
(2\pi\mu)^{-2\ve}\delta^d\left(p_A^\mu+p_B^\mu-\sum_{i=1}^N p_i^\mu\right)
\\ \nn && \qquad\qquad\qquad \times
\Psi^{(2N)}(\vec{p}) 
\frac{\as}{2\pi} \frac{(4\pi)^\ve}{\ve\Gamma(1-\ve)}
\left(\frac{\mu^2}{x_A s\delta}\right)^\ve
\int \d z(1-z)^{-2\ve}
\\ \nn && \qquad\qquad\qquad \times
\Bigg[\frac{1}{z^2}\Theta(z>x_A)
\sum_b\frac{\omega(b)}{\omega(a_A)}
\P_{a_A/b}(z,\ve) L\left(b,a_B,\frac{x_A}{z},x_B\right)
\\ \nn && \qquad\qquad\qquad\qquad
-\Theta(z>1-\alpha)L(a_A,a_B,x_A,x_B)\frac{2C(a_A)}{1-z}\Bigg].
\eeqn
In order that we could combine this contribution with the collinear
factorization counter term for hadron $A$, we use the relations
\beq
\frac{(4\pi)^\ve}{\ve\Gamma(1-\ve)}
=\cG\left(\frac{\mu^2}{\qes}\right)^\ve
\left[\frac{1}{\ve}+\ln\left(\frac{\qes}{\mu^2}\right)
+\O\,(\ve)\right]
\eeq
and
\beqn
&&X^\ve \int_0^1 \d z (1-z)^{-2\ve}
\left[f(z) P(z,\ve)\Theta(z>x)-f(1)\frac{2C}{1-z}\Theta(z>1-\alpha)\right]
\\ \nn && \qquad
= \int_{x}^1\d z f(z)\left[P(z,0)-\frac{2C}{1-z}+\frac{2C}{[1-z]_+}
+\gamma\delta(1-z)\right] - f(1)[\gamma+2 C \ln \alpha]
\\ \nn && \qquad
+\ve \Bigg\{\int_x^1 \d z \left[\ln\frac{X}{(1-z)^2}
\left(f(z) P(z,\ve)-f(1)\frac{2C}{1-z}\right) + f(z)P'(z)\right]
\\ \nn && \qquad\quad
+2Cf(1)\left[\ln X(\ln(1-x)-\ln\alpha)+\ln^2\alpha-\ln^2(1-x)\right]\Bigg\}
+\O\,(\ve)
\eeqn
The latter equation holds if $(1-z)P(z,0)\to 2C$ as $z\to 1$ and for any
function $f(z)$ that is not singular in $z=1$. The function $\P'(z)$ is
defined by the relation $\P(z,\ve)=\P(z)+\ve \P'(z)$. For the
complete collinear integral $\I[\coll]_A^A$ we obtain
\beqn
&&\I[\coll]_A^A=
\\ \nn &&\sum_{a_A,a_B,a_1,\ldots,a_N}
\int \D_N(\ve) \S_N(p_1^\mu,\ldots,p_N^\mu) 
(2\pi\mu)^{-2\ve}\delta^d\left(p_A^\mu+p_B^\mu-\sum_{i=1}^N p_i^\mu\right)
\\ \nn && \qquad\qquad\quad 
\times\Bigg\{
\frac{\as}{2\pi}\cG\left(\frac{\mu^2}{\qes}\right)^\ve
\Psi_A^\coll(\vec{a},\vec{p})L(a_A,a_B,x_A,x_B)
\\ \nn && \qquad\qquad\qquad
-\int_{x_A}^1\frac{\d z}{z^2} 
\sum_b\frac{\omega(b)}{\omega(a_A)}L(a_A,a_B,x_A/z,x_B)
\frac{(4\pi)^\ve}{\ve\Gamma(1-\ve)}
\frac{\as}{2\pi}P_{a_A/b}(z)\Psi^{(2N)}(\vec{a},\vec{p})
\\ \nn && \qquad\qquad\qquad
-\frac{\as}{2\pi}\cG\left(\frac{\mu^2}{\qes}\right)^\ve
\Psi^{(2N)}(\vec{a},\vec{p}) L(a_A,a_B,x_A,x_B)
\\ \nn && \qquad\qquad\qquad\quad\times
\Bigg[\int_{x_A}^1 \d z \ln X_A(z)
\Bigg(\frac{1}{z^2}\sum_b \frac{\omega(b)}{\omega(a_A)}
\frac{L\left(b,a_B,x_A/z,x_B\right)}{L(a_A,a_B,x_A,x_B)}\P_{a_A/b}(z)
-\frac{2C(a_A)}{1-z}\Bigg)
\\ \nn && \qquad\qquad\qquad\qquad
+\int_{x_A}^1 \d z \frac{1}{z^2}\sum_b \frac{\omega(b)}{\omega(a_A)}
\frac{L\left(b,a_B,x_A/z,x_B\right)}{L(a_A,a_B,x_A,x_B)}
\P'_{a_A/b}(z)
\\ \nn && \qquad\qquad\qquad\qquad
-2C(a_A)\left(\ln X_A(0) \ln\frac{\alpha}{1-x_A}
+\ln^2(1-x_A)-\ln^2\alpha\right)\Bigg]
\Bigg\}+\O\,(\ve),
\eeqn
where $X_A(z) = \mu^2/((1-z)^2 x_A s\delta)$ and
\beq
\Psi_A^\coll(\vec{a},\vec{p}) = \Psi^{(2N)}(\vec{a},\vec{p})
\Bigg(\frac{1}{\ve}[\gamma(a_A)+2 C(a_A) \ln \alpha]
+\ln\left(\frac{\qes}{\mu^2}\right)
[\gamma(a_A)+2 C(a_A) \ln \alpha]\Bigg).
\eeq

When two final state particles become collinear, $p_m^\mu=z p_P^\mu$,
$p^\mu=(1-z) p_P^\mu$, the measurement function
$\S_{N+1}(p_1^\mu,\ldots p^\mu)=\S_N(p_1^\mu,\ldots,p_N^\mu)\st{m\to P}$.
Keeping the helicity independent term in the collinear integral, we have
\beqn
&&\I[\coll]_m^m=
\\ \nn &&\sum_{{a_A,a_B,a_1,\ldots,a_{m-1}}\choose {a_P,\ldots,a_N}}
\int \Bigg[\D_N(\ve) \S_N(p_1^\mu,\ldots,p_N^\mu)
(2\pi\mu)^{-2\ve} 
\delta^d\left(p_A^\mu+p_B^\mu-\sum_{i=1}^N p_i^\mu\right)
\Bigg]\st{m\to P} 
\\ \nn && \qquad\qquad\qquad \times
\frac{\as}{(2\pi)^2} 
L(a_A,a_B,x_A,x_B)\left(\frac{2\pi\mu}{E_P}\right)^{2\ve}
\Psi^{(2N)}(\vec{a},\vec{p})\st{m\to P}
\\ \nn && \qquad\qquad\qquad \times
\int (1-z)^{-\ve}\d z
\left(\frac{Q^2}{E_P^2}\right)^{-\ve}\frac{\d Q^2}{Q^2}\,\d^{1-2\ve}\varphi
\\ \nn &&\qquad\qquad\qquad\qquad \times 
\Bigg[z^{-\ve}\,\Theta(4 z (1-z)E_P^2\delta>Q^2)\,\Theta(z > 1/2) 
\sum_{a_m}\tilde{P}_{a_m/a_P}(z,\ve)
\\ \nn &&\qquad\qquad\qquad\qquad \quad
-\Theta(4 (1-z)E_P^2\delta>Q^2)\,\Theta(z>1-\alpha)\,\frac{2C(a_m)}{1-z}
\Bigg].
\eeqn
The $Q^2$, $\varphi$ integrations can be calculated immediately. The
required integral is given in the appendix, eq.\ (\ref{Q2phi}).
The remaining integral over $z$ is also straightforward.
After performing these steps and leaving out 
the formal $m\to P$ substitutions, we see that the term $\I[\coll]_m^m$ 
has a form very similar to that of $\I[2\to N]$:
\beqn
&&\I[\coll]_m^m=
\\ \nn &&\sum_{a_A,a_B,a_1,\ldots,a_N}
\int \D_N(\ve) \S_N(p_1^\mu,\ldots,p_N^\mu)(2\pi\mu)^{-2\ve} 
\delta^d\left(p_A^\mu+p_B^\mu-\sum_{i=1}^N p_i^\mu\right)
\\ \nn &&\qquad\qquad\qquad \times 
L(a_A,a_B,x_A,x_B)\frac{\as}{2\pi} \cG
\left(\frac{\mu^2}{\qes}\right)^\ve \Psi^{\coll}_m(\vec{a},\vec{p}),
\eeqn
where
\beqn
&&\Psi^{\coll}_m(\vec{a},\vec{p}) =-\frac{1}{\ve}
\left(\frac{\qes}{4 E_m^2\delta}\right)^\ve
\Z_{a_m}(\alpha) \Psi^{(2N)}(\vec{a},\vec{p})+\O\,(\ve)
\eeqn
with $\Z_{a_m}(\alpha)$ given in eq.\ (\ref{Zam}).

At this point we see that the sum of the integrals $\I[2\to N]$,
$\I[\soft]$ and $\I[\coll]$ is free of any poles of 
$\ve$, therefore, it can be calculated in $d=4$ dimensions:
\beqn
&&\left(\I[2\to N]+\I[\soft]+\I[\coll]\right)\st{\ve=0}=
\\ \nn && \sum_{a_A,a_B,a_1,\ldots,a_N}
\int \D_N(0)\S_N(p_1^\mu,\ldots,p_N^\mu)
\delta^{(4)}\left(p_A^\mu+p_B^\mu-\sum_{i=1}^N p_i^\mu\right)
L(a_A,a_B,x_A,x_B)\Psi^{(2N)}(\vec{a},\vec{p})
\\ \nn &&\qquad\;\times
\Bigg\{1+\frac{\as}{2\pi}\Bigg(
\frac{\Psi^{(2N+2)}_{{\rm NS}}(\vec{a},\vec{p})}{\Psi^{(2N)}(\vec{a},\vec{p})}
+\sum_{{m,n=A,B,1,\ldots,N} \choose {m\ne n}} \tilde{\J}_{mn}(\vec{p})
\frac{\Psi_{mn}^{(2N;c)}(\vec{a},\vec{p})}{\Psi^{(2N)}(\vec{a},\vec{p})}
\\ \nn &&\qquad\qquad\qquad\quad
+\sum_{m=1}^N\Big[\gamma(a_m)+2C(a_m)\ln \alpha\Big]
\ln\frac{\qes}{4E_m^2\delta}
\\ \nn &&\qquad\qquad\qquad\quad
+\sum_{m=A,B}\Big[\gamma(a_m)+2C(a_m)\ln \alpha\Big]
\ln\frac{\qes}{\mu^2}
\\ \nn &&\qquad\qquad\qquad\quad
-\sum_{m=1}^N\left[2C(a_m)\left(\ln^2 \alpha+\frac{\pi^2}{3}\right)
+\gamma'(a_m)\right]
\\ \nn &&\qquad\qquad\qquad\quad
+\sum_{m=A,B}
2 C(a_m)\left[\ln X_m(0)
\ln\frac{\alpha}{1-x_m}+\ln^2(1-x_m)-\ln^2 \alpha\right]
\\ \nn &&\qquad\qquad\qquad\quad
-\int_{x_A}^1 \d z \ln X_A(z)
\Bigg[\frac{1}{z^2}\sum_b\frac{L(b,a_B,x_A/z,x_B)}{L(a_A,a_B,x_A,x_B)}
\P_{a_A/b}(z) -\frac{2C(a_B)}{1-z}\Bigg]
\\ \nn &&\qquad\qquad\qquad\quad
-\int_{x_B}^1 \d z \ln X_B(z)
\Bigg[\frac{1}{z^2}\sum_b\frac{L(a_B,b,x_A,x_B/z)}{L(a_A,a_B,x_A,x_B)}
\P_{a_B/b}(z) -\frac{2C(a_B)}{1-z}\Bigg]
\\ \nn && \qquad\qquad\qquad\quad
-\int_{x_A}^1 \d z \frac{1}{z^2}
\sum_b\frac{L(b,a_B,x_A/z,x_B)}{L(a_A,a_B,x_A,x_B)} \P'_{a_A/b}(z)
\\ \nn && \qquad\qquad\qquad\quad
-\int_{x_B}^1 \d z \frac{1}{z^2}
\sum_b\frac{L(a_B,b,x_A,x_B/z)}{L(a_A,a_B,x_A,x_B)} \P'_{a_B/b}(z)
\Bigg) \Bigg\},
\eeqn
The logarithmic dependence on the unphysical parameters $\alpha$ and
$\delta$ in this equation gets canceled when 
this contribution is combined with the integral $\I[\fin]$ (eq.\
(\ref{Ifin})) in order to obtain the infrared safe physical cross section
at \NLO:
\beq
\sigma = \I[\fin]\st{\ve = 0} + (\I[2\to N] + \I[\soft] + \I[\coll] )
\st{\ve = 0}.
\eeq

We can also demonstrate the independence of the $N$-body integral
of the auxiliary parameter $\qes$ explicitly by making use of
eqs. (\ref{Psi2N+2ns}), (\ref{softcollidentity}), (\ref{Itilde}) as
well as the definition of the $\ell_2$ function, formula (\ref{elldef}):
\beqn
\label{INbody}
&&\left(\I[2\to N]+\I[\soft]+\I[\coll]\right)\st{\ve=0}=
\\ \nn && \sum_{a_A,a_B,a_1,\ldots,a_N}
\int \D_N(0)\S_N(p_1^\mu,\ldots,p_N^\mu)
\delta^{(4)}\left(p_A^\mu+p_B^\mu-\sum_{i=1}^N p_i^\mu\right)
L(a_A,a_B,x_A,x_B)\Psi^{(2N)}(\vec{a},\vec{p})
\\ \nn &&\qquad\;\times
\Bigg\{1+\frac{\as}{2\pi}\Bigg(
\left[g^{2N}\left(\frac{g}{4\pi}\right)^2\right]^{-1}
\frac{1}{2} \sum_{{\rm hel}} \sum_{{\rm col}}
\left(\Am^{(1)}_{{\rm NS}}\Am^{(0)*}
+\Am^{(1)*}_{{\rm NS}}\Am^{(0)}\right)/\Psi^{(2N)}(\vec{a},\vec{p})
\\ \nn &&\qquad\qquad\qquad\quad
+\sum_{m=1}^N\sum_{{n=A,B,1,\ldots,N} \choose {m\ne n}}
\left[\frac{\pi^2}{4}
\Theta(\s mn)+\ln \alpha\ln\frac{\s mn}{4E_m^2\delta}\right]
\frac{\Psi_{mn}^{(2N;c)}(\vec{a},\vec{p})}{\Psi^{(2N)}(\vec{a},\vec{p})}
\\ \nn &&\qquad\qquad\qquad\quad
+\sum_{m=A,B}\sum_{{n=A,B,1,\ldots,N} \choose {m\ne n}}
\left[\frac{\pi^2}{4} \Theta(\s mn)
+\ln \alpha\ln\frac{\s mn}{x_m s\delta}\right]
\frac{\Psi_{mn}^{(2N;c)}(\vec{a},\vec{p})}{\Psi^{(2N)}(\vec{a},\vec{p})}
\\ \nn &&\qquad\qquad\qquad\quad
+\sum_{m=1}^N \Bigg[\gamma(a_m)\ln\frac{\mu^2}{4E_m^2\delta}
+\frac{\Nc}{6}-C(a_m)\frac{5\pi^2}{6}-\gamma'(a_m)
-\tilde{\gamma}(a_m)\Bigg]
\\ \nn &&\qquad\qquad\qquad\quad
+ \sum_{m=A,B}\Bigg[2C(a_m)\bigg(\!\ln(1\!-\!x_m)
\ln\frac{x_m (1\!-\!x_m)s\delta}{\mu^2}-\frac{\pi^2}{12}\bigg)
-\tilde{\gamma}(a_m)\Bigg]
\\ \nn &&\qquad\qquad\qquad\quad
-\int_{x_A}^1 \d z \ln X_A(z)
\Bigg[\frac{1}{z^2}\sum_b\frac{L(b,a_B,x_A/z,x_B)}{L(a_A,a_B,x_A,x_B)}
\P_{a_A/b}(z) -\frac{2C(a_A)}{1-z}\Bigg]
\\ \nn &&\qquad\qquad\qquad\quad
-\int_{x_B}^1 \d z \ln X_B(z)
\Bigg[\frac{1}{z^2}\sum_b\frac{L(a_B,b,x_A,x_B/z)}{L(a_A,a_B,x_A,x_B)}
\P_{a_B/b}(z) -\frac{2C(a_B)}{1-z}\Bigg]
\\ \nn && \qquad\qquad\qquad\quad
-\int_{x_A}^1 \d z \frac{1}{z^2}
\sum_b\frac{L(b,a_B,x_A/z,x_B)}{L(a_A,a_B,x_A,x_B)} \P'_{a_A/b}(z)
\\ \nn && \qquad\qquad\qquad\quad
-\int_{x_B}^1 \d z \frac{1}{z^2}
\sum_b\frac{L(a_B,b,x_A,x_B/z)}{L(a_A,a_B,x_A,x_B)} \P'_{a_B/b}(z)
\Bigg) \Bigg\}.
\eeqn
This equation together with eqs.\ (\ref{IfinA}) and (\ref{Ifinf})
define explicitly those integrals that are needed for the calculation
of a jet cross section at \NLO in perturbative QCD.

\section{Numerical results}
\setcounter{equation}{0}

In this section we present some numerical results of the first
non-trivial application of our algorithm, namely the calculation of
three-jet cross sections in $e^+e^-$ annihilation. Thus our results can
be compared with those of ref.\ \cite{KN3jet}.

We use the matrix elements of ref.\ \cite{ERT}  for the construction of
the various $\Psi$ functions. The algorithm can easily be altered for
performing jet cross section calculations in the case of $e^+e^-$
annihilation. One simply drops all terms in the integrals (\ref{Ifinf})
and (\ref{INbody}) that carry $A$ or $B$ indices, leaves out the $x_A$,
$x_B$ integrations from $\D_N(0)$  and then the sum of integrals
(\ref{Ifinf}) and (\ref{INbody}) immediately gives the physical cross
section. We implemented such an algorithm in a Monte Carlo program.
The results are in good agreement with those of ref.\ \cite{KN3jet}.
As an example we show the \NLO coefficients for the thrust, C-parameter
distributions multiplied by $(1-t)$ and $C$ respectively and
distributions for the Jade E and $k_\perp$ jet clustering algorithms
multiplied by the jet resolution parameter in fig.\ 1. We find that
the numerical convergence is similar to the program of ref.\ \cite{KN3jet}.

\newcount\fc \fc=0
\def\fig{\global\advance\fc by 1 Fig.\ \the\fc: }

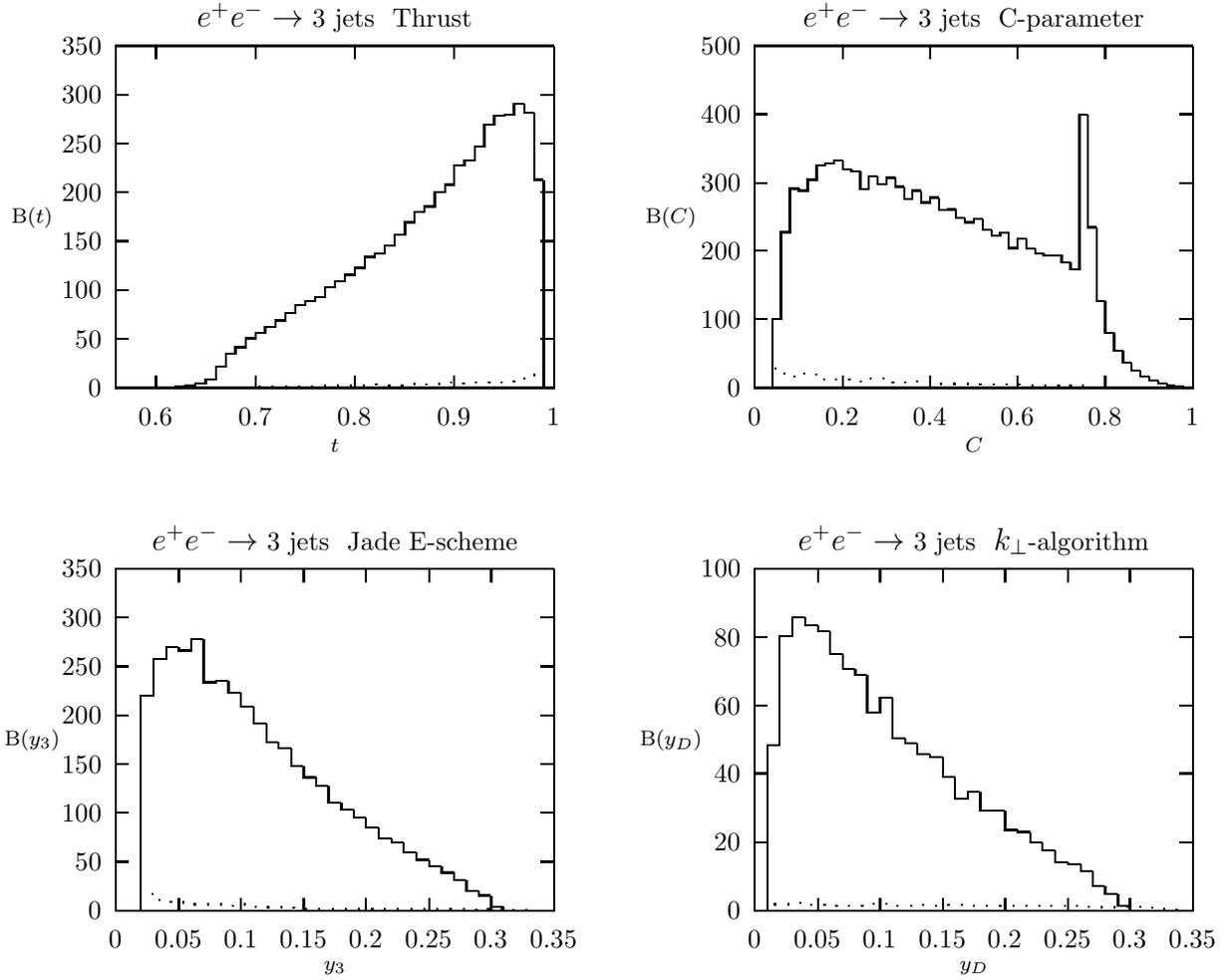
\begin{figure}
%\hbox{\input{thrustplot.tex} \input{cparplot.tex}}
\hbox{
% GNUPLOT: LaTeX picture
\setlength{\unitlength}{0.240900pt}
\ifx\plotpoint\undefined\newsavebox{\plotpoint}\fi
\begin{picture}(974,720)(0,0)
\font\gnuplot=cmr10 at 10pt
\gnuplot
\sbox{\plotpoint}{\rule[-0.200pt]{0.400pt}{0.400pt}}%
\put(220.0,113.0){\rule[-0.200pt]{166.221pt}{0.400pt}}
\put(220.0,113.0){\rule[-0.200pt]{4.818pt}{0.400pt}}
\put(198,113){\makebox(0,0)[r]{0}}
\put(890.0,113.0){\rule[-0.200pt]{4.818pt}{0.400pt}}
\put(220.0,190.0){\rule[-0.200pt]{4.818pt}{0.400pt}}
\put(198,190){\makebox(0,0)[r]{50}}
\put(890.0,190.0){\rule[-0.200pt]{4.818pt}{0.400pt}}
\put(220.0,267.0){\rule[-0.200pt]{4.818pt}{0.400pt}}
\put(198,267){\makebox(0,0)[r]{100}}
\put(890.0,267.0){\rule[-0.200pt]{4.818pt}{0.400pt}}
\put(220.0,344.0){\rule[-0.200pt]{4.818pt}{0.400pt}}
\put(198,344){\makebox(0,0)[r]{150}}
\put(890.0,344.0){\rule[-0.200pt]{4.818pt}{0.400pt}}
\put(220.0,421.0){\rule[-0.200pt]{4.818pt}{0.400pt}}
\put(198,421){\makebox(0,0)[r]{200}}
\put(890.0,421.0){\rule[-0.200pt]{4.818pt}{0.400pt}}
\put(220.0,498.0){\rule[-0.200pt]{4.818pt}{0.400pt}}
\put(198,498){\makebox(0,0)[r]{250}}
\put(890.0,498.0){\rule[-0.200pt]{4.818pt}{0.400pt}}
\put(220.0,575.0){\rule[-0.200pt]{4.818pt}{0.400pt}}
\put(198,575){\makebox(0,0)[r]{300}}
\put(890.0,575.0){\rule[-0.200pt]{4.818pt}{0.400pt}}
\put(220.0,652.0){\rule[-0.200pt]{4.818pt}{0.400pt}}
\put(198,652){\makebox(0,0)[r]{350}}
\put(890.0,652.0){\rule[-0.200pt]{4.818pt}{0.400pt}}
\put(283.0,113.0){\rule[-0.200pt]{0.400pt}{4.818pt}}
\put(283,68){\makebox(0,0){0.6}}
\put(283.0,632.0){\rule[-0.200pt]{0.400pt}{4.818pt}}
\put(440.0,113.0){\rule[-0.200pt]{0.400pt}{4.818pt}}
\put(440,68){\makebox(0,0){0.7}}
\put(440.0,632.0){\rule[-0.200pt]{0.400pt}{4.818pt}}
\put(596.0,113.0){\rule[-0.200pt]{0.400pt}{4.818pt}}
\put(596,68){\makebox(0,0){0.8}}
\put(596.0,632.0){\rule[-0.200pt]{0.400pt}{4.818pt}}
\put(753.0,113.0){\rule[-0.200pt]{0.400pt}{4.818pt}}
\put(753,68){\makebox(0,0){0.9}}
\put(753.0,632.0){\rule[-0.200pt]{0.400pt}{4.818pt}}
\put(910.0,113.0){\rule[-0.200pt]{0.400pt}{4.818pt}}
\put(910,68){\makebox(0,0){1}}
\put(910.0,632.0){\rule[-0.200pt]{0.400pt}{4.818pt}}
\put(220.0,113.0){\rule[-0.200pt]{166.221pt}{0.400pt}}
\put(910.0,113.0){\rule[-0.200pt]{0.400pt}{129.845pt}}
\put(220.0,652.0){\rule[-0.200pt]{166.221pt}{0.400pt}}
\put(89,382){\makebox(0,0){\scriptsize B($t$)}}
\put(565,23){\makebox(0,0){ \scriptsize $t$ }}
\put(565,697){\makebox(0,0){$e^+e^- \to$ 3 jets\,\, Thrust}}
\put(220.0,113.0){\rule[-0.200pt]{0.400pt}{129.845pt}}
\put(220,113){\usebox{\plotpoint}}
\put(220,113){\usebox{\plotpoint}}
\put(220.0,113.0){\rule[-0.200pt]{18.790pt}{0.400pt}}
\put(298.0,113.0){\usebox{\plotpoint}}
\put(298.0,114.0){\rule[-0.200pt]{3.854pt}{0.400pt}}
\put(314.0,114.0){\usebox{\plotpoint}}
\put(314.0,115.0){\rule[-0.200pt]{3.854pt}{0.400pt}}
\put(330.0,115.0){\rule[-0.200pt]{0.400pt}{0.482pt}}
\put(330.0,117.0){\rule[-0.200pt]{3.613pt}{0.400pt}}
\put(345.0,117.0){\rule[-0.200pt]{0.400pt}{0.723pt}}
\put(345.0,120.0){\rule[-0.200pt]{3.854pt}{0.400pt}}
\put(361.0,120.0){\rule[-0.200pt]{0.400pt}{1.445pt}}
\put(361.0,126.0){\rule[-0.200pt]{3.854pt}{0.400pt}}
\put(377.0,126.0){\rule[-0.200pt]{0.400pt}{4.818pt}}
\put(377.0,146.0){\rule[-0.200pt]{3.854pt}{0.400pt}}
\put(393.0,146.0){\rule[-0.200pt]{0.400pt}{5.059pt}}
\put(393.0,167.0){\rule[-0.200pt]{3.613pt}{0.400pt}}
\put(408.0,167.0){\rule[-0.200pt]{0.400pt}{2.409pt}}
\put(408.0,177.0){\rule[-0.200pt]{3.854pt}{0.400pt}}
\put(424.0,177.0){\rule[-0.200pt]{0.400pt}{3.373pt}}
\put(424.0,191.0){\rule[-0.200pt]{3.854pt}{0.400pt}}
\put(440.0,191.0){\rule[-0.200pt]{0.400pt}{2.168pt}}
\put(440.0,200.0){\rule[-0.200pt]{3.613pt}{0.400pt}}
\put(455.0,200.0){\rule[-0.200pt]{0.400pt}{2.168pt}}
\put(455.0,209.0){\rule[-0.200pt]{3.854pt}{0.400pt}}
\put(471.0,209.0){\rule[-0.200pt]{0.400pt}{2.409pt}}
\put(471.0,219.0){\rule[-0.200pt]{3.854pt}{0.400pt}}
\put(487.0,219.0){\rule[-0.200pt]{0.400pt}{2.891pt}}
\put(487.0,231.0){\rule[-0.200pt]{3.613pt}{0.400pt}}
\put(502.0,231.0){\rule[-0.200pt]{0.400pt}{2.891pt}}
\put(502.0,243.0){\rule[-0.200pt]{3.854pt}{0.400pt}}
\put(518.0,243.0){\rule[-0.200pt]{0.400pt}{1.686pt}}
\put(518.0,250.0){\rule[-0.200pt]{3.854pt}{0.400pt}}
\put(534.0,250.0){\rule[-0.200pt]{0.400pt}{1.445pt}}
\put(534.0,256.0){\rule[-0.200pt]{3.613pt}{0.400pt}}
\put(549.0,256.0){\rule[-0.200pt]{0.400pt}{3.613pt}}
\put(549.0,271.0){\rule[-0.200pt]{3.854pt}{0.400pt}}
\put(565.0,271.0){\rule[-0.200pt]{0.400pt}{2.409pt}}
\put(565.0,281.0){\rule[-0.200pt]{3.854pt}{0.400pt}}
\put(581.0,281.0){\rule[-0.200pt]{0.400pt}{2.409pt}}
\put(581.0,291.0){\rule[-0.200pt]{3.613pt}{0.400pt}}
\put(596.0,291.0){\rule[-0.200pt]{0.400pt}{2.650pt}}
\put(596.0,302.0){\rule[-0.200pt]{3.854pt}{0.400pt}}
\put(612.0,302.0){\rule[-0.200pt]{0.400pt}{4.095pt}}
\put(612.0,319.0){\rule[-0.200pt]{3.854pt}{0.400pt}}
\put(628.0,319.0){\rule[-0.200pt]{0.400pt}{1.445pt}}
\put(628.0,325.0){\rule[-0.200pt]{3.613pt}{0.400pt}}
\put(643.0,325.0){\rule[-0.200pt]{0.400pt}{2.891pt}}
\put(643.0,337.0){\rule[-0.200pt]{3.854pt}{0.400pt}}
\put(659.0,337.0){\rule[-0.200pt]{0.400pt}{4.095pt}}
\put(659.0,354.0){\rule[-0.200pt]{3.854pt}{0.400pt}}
\put(675.0,354.0){\rule[-0.200pt]{0.400pt}{4.818pt}}
\put(675.0,374.0){\rule[-0.200pt]{3.613pt}{0.400pt}}
\put(690.0,374.0){\rule[-0.200pt]{0.400pt}{3.854pt}}
\put(690.0,390.0){\rule[-0.200pt]{3.854pt}{0.400pt}}
\put(706.0,390.0){\rule[-0.200pt]{0.400pt}{2.168pt}}
\put(706.0,399.0){\rule[-0.200pt]{3.854pt}{0.400pt}}
\put(722.0,399.0){\rule[-0.200pt]{0.400pt}{5.300pt}}
\put(722.0,421.0){\rule[-0.200pt]{3.854pt}{0.400pt}}
\put(738.0,421.0){\rule[-0.200pt]{0.400pt}{2.891pt}}
\put(738.0,433.0){\rule[-0.200pt]{3.613pt}{0.400pt}}
\put(753.0,433.0){\rule[-0.200pt]{0.400pt}{7.468pt}}
\put(753.0,464.0){\rule[-0.200pt]{3.854pt}{0.400pt}}
\put(769.0,464.0){\rule[-0.200pt]{0.400pt}{1.686pt}}
\put(769.0,471.0){\rule[-0.200pt]{3.854pt}{0.400pt}}
\put(785.0,471.0){\rule[-0.200pt]{0.400pt}{5.300pt}}
\put(785.0,493.0){\rule[-0.200pt]{3.613pt}{0.400pt}}
\put(800.0,493.0){\rule[-0.200pt]{0.400pt}{8.431pt}}
\put(800.0,528.0){\rule[-0.200pt]{3.854pt}{0.400pt}}
\put(816.0,528.0){\rule[-0.200pt]{0.400pt}{3.373pt}}
\put(816.0,542.0){\rule[-0.200pt]{3.854pt}{0.400pt}}
\put(832.0,542.0){\usebox{\plotpoint}}
\put(832.0,543.0){\rule[-0.200pt]{3.613pt}{0.400pt}}
\put(847.0,543.0){\rule[-0.200pt]{0.400pt}{4.095pt}}
\put(847.0,560.0){\rule[-0.200pt]{3.854pt}{0.400pt}}
\put(863.0,546.0){\rule[-0.200pt]{0.400pt}{3.373pt}}
\put(863.0,546.0){\rule[-0.200pt]{3.854pt}{0.400pt}}
\put(879.0,441.0){\rule[-0.200pt]{0.400pt}{25.294pt}}
\put(879.0,441.0){\rule[-0.200pt]{3.613pt}{0.400pt}}
\put(894.0,113.0){\rule[-0.200pt]{0.400pt}{79.015pt}}
\put(220,113){\usebox{\plotpoint}}
\put(220.00,113.00){\usebox{\plotpoint}}
\put(240.76,113.00){\usebox{\plotpoint}}
\put(261.51,113.00){\usebox{\plotpoint}}
\put(282.27,113.00){\usebox{\plotpoint}}
\multiput(283,113)(20.756,0.000){0}{\usebox{\plotpoint}}
\put(303.02,113.00){\usebox{\plotpoint}}
\put(323.78,113.00){\usebox{\plotpoint}}
\put(344.53,113.00){\usebox{\plotpoint}}
\multiput(345,113)(20.756,0.000){0}{\usebox{\plotpoint}}
\put(365.29,113.00){\usebox{\plotpoint}}
\put(386.04,113.00){\usebox{\plotpoint}}
\multiput(393,113)(0.000,20.756){0}{\usebox{\plotpoint}}
\put(405.80,114.00){\usebox{\plotpoint}}
\multiput(408,114)(20.756,0.000){0}{\usebox{\plotpoint}}
\put(426.55,114.00){\usebox{\plotpoint}}
\multiput(440,114)(0.000,20.756){0}{\usebox{\plotpoint}}
\put(446.31,115.00){\usebox{\plotpoint}}
\put(467.07,115.00){\usebox{\plotpoint}}
\multiput(471,115)(20.756,0.000){0}{\usebox{\plotpoint}}
\put(487.82,115.00){\usebox{\plotpoint}}
\put(508.58,115.00){\usebox{\plotpoint}}
\put(529.33,115.00){\usebox{\plotpoint}}
\multiput(534,115)(0.000,20.756){0}{\usebox{\plotpoint}}
\multiput(534,116)(20.756,0.000){0}{\usebox{\plotpoint}}
\put(549.09,116.00){\usebox{\plotpoint}}
\put(569.84,116.00){\usebox{\plotpoint}}
\put(590.60,116.00){\usebox{\plotpoint}}
\multiput(596,116)(0.000,20.756){0}{\usebox{\plotpoint}}
\put(610.35,117.00){\usebox{\plotpoint}}
\multiput(612,117)(20.756,0.000){0}{\usebox{\plotpoint}}
\multiput(628,117)(0.000,20.756){0}{\usebox{\plotpoint}}
\put(630.11,118.00){\usebox{\plotpoint}}
\multiput(643,118)(0.000,-20.756){0}{\usebox{\plotpoint}}
\put(649.87,117.00){\usebox{\plotpoint}}
\put(670.62,117.00){\usebox{\plotpoint}}
\multiput(675,117)(0.000,20.756){0}{\usebox{\plotpoint}}
\put(689.38,119.00){\usebox{\plotpoint}}
\multiput(690,119)(20.756,0.000){0}{\usebox{\plotpoint}}
\multiput(706,119)(0.000,-20.756){0}{\usebox{\plotpoint}}
\put(709.13,118.00){\usebox{\plotpoint}}
\multiput(722,118)(0.000,20.756){0}{\usebox{\plotpoint}}
\put(728.89,119.00){\usebox{\plotpoint}}
\multiput(738,119)(0.000,20.756){0}{\usebox{\plotpoint}}
\put(747.64,121.00){\usebox{\plotpoint}}
\multiput(753,121)(0.000,-20.756){0}{\usebox{\plotpoint}}
\put(767.40,120.00){\usebox{\plotpoint}}
\multiput(769,120)(0.000,20.756){0}{\usebox{\plotpoint}}
\multiput(769,121)(20.756,0.000){0}{\usebox{\plotpoint}}
\put(787.15,121.00){\usebox{\plotpoint}}
\put(807.91,121.00){\usebox{\plotpoint}}
\multiput(816,121)(0.000,20.756){0}{\usebox{\plotpoint}}
\put(827.66,122.00){\usebox{\plotpoint}}
\multiput(832,122)(0.000,20.756){0}{\usebox{\plotpoint}}
\multiput(832,123)(20.756,0.000){0}{\usebox{\plotpoint}}
\put(847.00,123.42){\usebox{\plotpoint}}
\multiput(847,124)(20.756,0.000){0}{\usebox{\plotpoint}}
\multiput(863,124)(0.000,20.756){0}{\usebox{\plotpoint}}
\put(863.18,128.00){\usebox{\plotpoint}}
\put(879.00,132.93){\usebox{\plotpoint}}
\multiput(879,136)(20.756,0.000){0}{\usebox{\plotpoint}}
\put(894.00,133.31){\usebox{\plotpoint}}
\put(894,113){\usebox{\plotpoint}}
\end{picture}
% GNUPLOT: LaTeX picture
\setlength{\unitlength}{0.240900pt}
\ifx\plotpoint\undefined\newsavebox{\plotpoint}\fi
\sbox{\plotpoint}{\rule[-0.200pt]{0.400pt}{0.400pt}}%
\begin{picture}(974,720)(0,0)
\font\gnuplot=cmr10 at 10pt
\gnuplot
\sbox{\plotpoint}{\rule[-0.200pt]{0.400pt}{0.400pt}}%
\put(220.0,113.0){\rule[-0.200pt]{166.221pt}{0.400pt}}
\put(220.0,113.0){\rule[-0.200pt]{0.400pt}{129.845pt}}
\put(220.0,113.0){\rule[-0.200pt]{4.818pt}{0.400pt}}
\put(198,113){\makebox(0,0)[r]{0}}
\put(890.0,113.0){\rule[-0.200pt]{4.818pt}{0.400pt}}
\put(220.0,221.0){\rule[-0.200pt]{4.818pt}{0.400pt}}
\put(198,221){\makebox(0,0)[r]{100}}
\put(890.0,221.0){\rule[-0.200pt]{4.818pt}{0.400pt}}
\put(220.0,329.0){\rule[-0.200pt]{4.818pt}{0.400pt}}
\put(198,329){\makebox(0,0)[r]{200}}
\put(890.0,329.0){\rule[-0.200pt]{4.818pt}{0.400pt}}
\put(220.0,436.0){\rule[-0.200pt]{4.818pt}{0.400pt}}
\put(198,436){\makebox(0,0)[r]{300}}
\put(890.0,436.0){\rule[-0.200pt]{4.818pt}{0.400pt}}
\put(220.0,544.0){\rule[-0.200pt]{4.818pt}{0.400pt}}
\put(198,544){\makebox(0,0)[r]{400}}
\put(890.0,544.0){\rule[-0.200pt]{4.818pt}{0.400pt}}
\put(220.0,652.0){\rule[-0.200pt]{4.818pt}{0.400pt}}
\put(198,652){\makebox(0,0)[r]{500}}
\put(890.0,652.0){\rule[-0.200pt]{4.818pt}{0.400pt}}
\put(220.0,113.0){\rule[-0.200pt]{0.400pt}{4.818pt}}
\put(220,68){\makebox(0,0){0}}
\put(220.0,632.0){\rule[-0.200pt]{0.400pt}{4.818pt}}
\put(358.0,113.0){\rule[-0.200pt]{0.400pt}{4.818pt}}
\put(358,68){\makebox(0,0){0.2}}
\put(358.0,632.0){\rule[-0.200pt]{0.400pt}{4.818pt}}
\put(496.0,113.0){\rule[-0.200pt]{0.400pt}{4.818pt}}
\put(496,68){\makebox(0,0){0.4}}
\put(496.0,632.0){\rule[-0.200pt]{0.400pt}{4.818pt}}
\put(634.0,113.0){\rule[-0.200pt]{0.400pt}{4.818pt}}
\put(634,68){\makebox(0,0){0.6}}
\put(634.0,632.0){\rule[-0.200pt]{0.400pt}{4.818pt}}
\put(772.0,113.0){\rule[-0.200pt]{0.400pt}{4.818pt}}
\put(772,68){\makebox(0,0){0.8}}
\put(772.0,632.0){\rule[-0.200pt]{0.400pt}{4.818pt}}
\put(910.0,113.0){\rule[-0.200pt]{0.400pt}{4.818pt}}
\put(910,68){\makebox(0,0){1}}
\put(910.0,632.0){\rule[-0.200pt]{0.400pt}{4.818pt}}
\put(220.0,113.0){\rule[-0.200pt]{166.221pt}{0.400pt}}
\put(910.0,113.0){\rule[-0.200pt]{0.400pt}{129.845pt}}
\put(220.0,652.0){\rule[-0.200pt]{166.221pt}{0.400pt}}
\put(89,382){\makebox(0,0){\scriptsize B($C$)}}
\put(565,23){\makebox(0,0){ \scriptsize $C$ }}
\put(565,697){\makebox(0,0){$e^+e^- \to$ 3 jets\,\,  C-parameter}}
\put(220.0,113.0){\rule[-0.200pt]{0.400pt}{129.845pt}}
\put(220,113){\usebox{\plotpoint}}
\put(220.0,113.0){\rule[-0.200pt]{6.745pt}{0.400pt}}
\put(248.0,113.0){\rule[-0.200pt]{0.400pt}{26.258pt}}
\put(248.0,222.0){\rule[-0.200pt]{3.132pt}{0.400pt}}
\put(261.0,222.0){\rule[-0.200pt]{0.400pt}{32.762pt}}
\put(261.0,358.0){\rule[-0.200pt]{3.373pt}{0.400pt}}
\put(275.0,358.0){\rule[-0.200pt]{0.400pt}{16.622pt}}
\put(275.0,427.0){\rule[-0.200pt]{3.373pt}{0.400pt}}
\put(289.0,424.0){\rule[-0.200pt]{0.400pt}{0.723pt}}
\put(289.0,424.0){\rule[-0.200pt]{3.373pt}{0.400pt}}
\put(303.0,424.0){\rule[-0.200pt]{0.400pt}{4.095pt}}
\put(303.0,441.0){\rule[-0.200pt]{3.373pt}{0.400pt}}
\put(317.0,441.0){\rule[-0.200pt]{0.400pt}{5.541pt}}
\put(317.0,464.0){\rule[-0.200pt]{3.132pt}{0.400pt}}
\put(330.0,464.0){\rule[-0.200pt]{0.400pt}{0.723pt}}
\put(330.0,467.0){\rule[-0.200pt]{3.373pt}{0.400pt}}
\put(344.0,467.0){\rule[-0.200pt]{0.400pt}{0.964pt}}
\put(344.0,471.0){\rule[-0.200pt]{3.373pt}{0.400pt}}
\put(358.0,457.0){\rule[-0.200pt]{0.400pt}{3.373pt}}
\put(358.0,457.0){\rule[-0.200pt]{3.373pt}{0.400pt}}
\put(372.0,454.0){\rule[-0.200pt]{0.400pt}{0.723pt}}
\put(372.0,454.0){\rule[-0.200pt]{3.373pt}{0.400pt}}
\put(386.0,426.0){\rule[-0.200pt]{0.400pt}{6.745pt}}
\put(386.0,426.0){\rule[-0.200pt]{3.132pt}{0.400pt}}
\put(399.0,426.0){\rule[-0.200pt]{0.400pt}{4.818pt}}
\put(399.0,446.0){\rule[-0.200pt]{3.373pt}{0.400pt}}
\put(413.0,434.0){\rule[-0.200pt]{0.400pt}{2.891pt}}
\put(413.0,434.0){\rule[-0.200pt]{3.373pt}{0.400pt}}
\put(427.0,434.0){\rule[-0.200pt]{0.400pt}{2.409pt}}
\put(427.0,444.0){\rule[-0.200pt]{3.373pt}{0.400pt}}
\put(441.0,430.0){\rule[-0.200pt]{0.400pt}{3.373pt}}
\put(441.0,430.0){\rule[-0.200pt]{3.373pt}{0.400pt}}
\put(455.0,411.0){\rule[-0.200pt]{0.400pt}{4.577pt}}
\put(455.0,411.0){\rule[-0.200pt]{3.132pt}{0.400pt}}
\put(468.0,411.0){\rule[-0.200pt]{0.400pt}{3.132pt}}
\put(468.0,424.0){\rule[-0.200pt]{3.373pt}{0.400pt}}
\put(482.0,405.0){\rule[-0.200pt]{0.400pt}{4.577pt}}
\put(482.0,405.0){\rule[-0.200pt]{3.373pt}{0.400pt}}
\put(496.0,405.0){\rule[-0.200pt]{0.400pt}{1.927pt}}
\put(496.0,413.0){\rule[-0.200pt]{3.373pt}{0.400pt}}
\put(510.0,393.0){\rule[-0.200pt]{0.400pt}{4.818pt}}
\put(510.0,393.0){\rule[-0.200pt]{3.373pt}{0.400pt}}
\put(524.0,393.0){\usebox{\plotpoint}}
\put(524.0,394.0){\rule[-0.200pt]{3.132pt}{0.400pt}}
\put(537.0,381.0){\rule[-0.200pt]{0.400pt}{3.132pt}}
\put(537.0,381.0){\rule[-0.200pt]{3.373pt}{0.400pt}}
\put(551.0,374.0){\rule[-0.200pt]{0.400pt}{1.686pt}}
\put(551.0,374.0){\rule[-0.200pt]{3.373pt}{0.400pt}}
\put(565.0,374.0){\rule[-0.200pt]{0.400pt}{1.204pt}}
\put(565.0,379.0){\rule[-0.200pt]{3.373pt}{0.400pt}}
\put(579.0,362.0){\rule[-0.200pt]{0.400pt}{4.095pt}}
\put(579.0,362.0){\rule[-0.200pt]{3.373pt}{0.400pt}}
\put(593.0,353.0){\rule[-0.200pt]{0.400pt}{2.168pt}}
\put(593.0,353.0){\rule[-0.200pt]{3.132pt}{0.400pt}}
\put(606.0,353.0){\rule[-0.200pt]{0.400pt}{0.964pt}}
\put(606.0,357.0){\rule[-0.200pt]{3.373pt}{0.400pt}}
\put(620.0,333.0){\rule[-0.200pt]{0.400pt}{5.782pt}}
\put(620.0,333.0){\rule[-0.200pt]{3.373pt}{0.400pt}}
\put(634.0,333.0){\rule[-0.200pt]{0.400pt}{3.613pt}}
\put(634.0,348.0){\rule[-0.200pt]{3.373pt}{0.400pt}}
\put(648.0,332.0){\rule[-0.200pt]{0.400pt}{3.854pt}}
\put(648.0,332.0){\rule[-0.200pt]{3.373pt}{0.400pt}}
\put(662.0,325.0){\rule[-0.200pt]{0.400pt}{1.686pt}}
\put(662.0,325.0){\rule[-0.200pt]{3.132pt}{0.400pt}}
\put(675.0,321.0){\rule[-0.200pt]{0.400pt}{0.964pt}}
\put(675.0,321.0){\rule[-0.200pt]{3.373pt}{0.400pt}}
\put(689.0,321.0){\usebox{\plotpoint}}
\put(689.0,322.0){\rule[-0.200pt]{3.373pt}{0.400pt}}
\put(703.0,311.0){\rule[-0.200pt]{0.400pt}{2.650pt}}
\put(703.0,311.0){\rule[-0.200pt]{3.373pt}{0.400pt}}
\put(717.0,300.0){\rule[-0.200pt]{0.400pt}{2.650pt}}
\put(717.0,300.0){\rule[-0.200pt]{3.373pt}{0.400pt}}
\put(731.0,300.0){\rule[-0.200pt]{0.400pt}{58.539pt}}
\put(731.0,543.0){\rule[-0.200pt]{3.132pt}{0.400pt}}
\put(744.0,366.0){\rule[-0.200pt]{0.400pt}{42.639pt}}
\put(744.0,366.0){\rule[-0.200pt]{3.373pt}{0.400pt}}
\put(758.0,250.0){\rule[-0.200pt]{0.400pt}{27.944pt}}
\put(758.0,250.0){\rule[-0.200pt]{3.373pt}{0.400pt}}
\put(772.0,200.0){\rule[-0.200pt]{0.400pt}{12.045pt}}
\put(772.0,200.0){\rule[-0.200pt]{3.373pt}{0.400pt}}
\put(786.0,171.0){\rule[-0.200pt]{0.400pt}{6.986pt}}
\put(786.0,171.0){\rule[-0.200pt]{3.373pt}{0.400pt}}
\put(800.0,153.0){\rule[-0.200pt]{0.400pt}{4.336pt}}
\put(800.0,153.0){\rule[-0.200pt]{3.132pt}{0.400pt}}
\put(813.0,140.0){\rule[-0.200pt]{0.400pt}{3.132pt}}
\put(813.0,140.0){\rule[-0.200pt]{3.373pt}{0.400pt}}
\put(827.0,131.0){\rule[-0.200pt]{0.400pt}{2.168pt}}
\put(827.0,131.0){\rule[-0.200pt]{3.373pt}{0.400pt}}
\put(841.0,125.0){\rule[-0.200pt]{0.400pt}{1.445pt}}
\put(841.0,125.0){\rule[-0.200pt]{3.373pt}{0.400pt}}
\put(855.0,120.0){\rule[-0.200pt]{0.400pt}{1.204pt}}
\put(855.0,120.0){\rule[-0.200pt]{3.373pt}{0.400pt}}
\put(869.0,117.0){\rule[-0.200pt]{0.400pt}{0.723pt}}
\put(869.0,117.0){\rule[-0.200pt]{3.132pt}{0.400pt}}
\put(882.0,115.0){\rule[-0.200pt]{0.400pt}{0.482pt}}
\put(882.0,115.0){\rule[-0.200pt]{3.373pt}{0.400pt}}
\put(896.0,113.0){\rule[-0.200pt]{0.400pt}{0.482pt}}
\put(220,113){\usebox{\plotpoint}}
\put(220.00,113.00){\usebox{\plotpoint}}
\put(240.76,113.00){\usebox{\plotpoint}}
\put(248.00,126.51){\usebox{\plotpoint}}
\put(252.27,143.00){\usebox{\plotpoint}}
\multiput(261,143)(0.000,-20.756){0}{\usebox{\plotpoint}}
\put(266.02,136.00){\usebox{\plotpoint}}
\multiput(275,136)(0.000,-20.756){0}{\usebox{\plotpoint}}
\put(281.78,131.00){\usebox{\plotpoint}}
\multiput(289,131)(0.000,20.756){0}{\usebox{\plotpoint}}
\put(297.53,136.00){\usebox{\plotpoint}}
\multiput(303,136)(0.000,-20.756){0}{\usebox{\plotpoint}}
\put(316.29,134.00){\usebox{\plotpoint}}
\multiput(317,134)(0.000,-20.756){0}{\usebox{\plotpoint}}
\put(329.04,126.00){\usebox{\plotpoint}}
\multiput(330,126)(0.000,-20.756){0}{\usebox{\plotpoint}}
\multiput(330,125)(20.756,0.000){0}{\usebox{\plotpoint}}
\multiput(344,125)(0.000,20.756){0}{\usebox{\plotpoint}}
\put(347.80,126.00){\usebox{\plotpoint}}
\put(368.55,126.00){\usebox{\plotpoint}}
\multiput(372,126)(0.000,-20.756){0}{\usebox{\plotpoint}}
\multiput(372,123)(20.756,0.000){0}{\usebox{\plotpoint}}
\put(386.31,123.00){\usebox{\plotpoint}}
\multiput(399,123)(0.000,20.756){0}{\usebox{\plotpoint}}
\put(402.07,128.00){\usebox{\plotpoint}}
\put(422.82,128.00){\usebox{\plotpoint}}
\multiput(427,128)(0.000,-20.756){0}{\usebox{\plotpoint}}
\put(437.58,122.00){\usebox{\plotpoint}}
\multiput(441,122)(0.000,-20.756){0}{\usebox{\plotpoint}}
\multiput(441,121)(20.756,0.000){0}{\usebox{\plotpoint}}
\put(457.33,121.00){\usebox{\plotpoint}}
\multiput(468,121)(0.000,20.756){0}{\usebox{\plotpoint}}
\put(476.09,123.00){\usebox{\plotpoint}}
\multiput(482,123)(0.000,-20.756){0}{\usebox{\plotpoint}}
\put(495.84,122.00){\usebox{\plotpoint}}
\multiput(496,122)(0.000,-20.756){0}{\usebox{\plotpoint}}
\multiput(496,121)(20.756,0.000){0}{\usebox{\plotpoint}}
\multiput(510,121)(0.000,-20.756){0}{\usebox{\plotpoint}}
\put(513.60,119.00){\usebox{\plotpoint}}
\put(534.35,119.00){\usebox{\plotpoint}}
\multiput(537,119)(20.756,0.000){0}{\usebox{\plotpoint}}
\put(555.11,119.00){\usebox{\plotpoint}}
\multiput(565,119)(0.000,-20.756){0}{\usebox{\plotpoint}}
\put(574.87,118.00){\usebox{\plotpoint}}
\multiput(579,118)(20.756,0.000){0}{\usebox{\plotpoint}}
\put(595.62,118.00){\usebox{\plotpoint}}
\put(616.38,118.00){\usebox{\plotpoint}}
\multiput(620,118)(20.756,0.000){0}{\usebox{\plotpoint}}
\put(637.13,118.00){\usebox{\plotpoint}}
\multiput(648,118)(0.000,-20.756){0}{\usebox{\plotpoint}}
\put(656.89,117.00){\usebox{\plotpoint}}
\multiput(662,117)(20.756,0.000){0}{\usebox{\plotpoint}}
\put(677.64,117.00){\usebox{\plotpoint}}
\put(698.40,117.00){\usebox{\plotpoint}}
\multiput(703,117)(0.000,-20.756){0}{\usebox{\plotpoint}}
\multiput(703,116)(20.756,0.000){0}{\usebox{\plotpoint}}
\multiput(717,116)(0.000,20.756){0}{\usebox{\plotpoint}}
\put(717.15,117.00){\usebox{\plotpoint}}
\multiput(731,117)(0.000,-20.756){0}{\usebox{\plotpoint}}
\put(736.91,116.00){\usebox{\plotpoint}}
\multiput(744,116)(0.000,-20.756){0}{\usebox{\plotpoint}}
\put(755.66,114.00){\usebox{\plotpoint}}
\multiput(758,114)(0.000,-20.756){0}{\usebox{\plotpoint}}
\multiput(758,113)(20.756,0.000){0}{\usebox{\plotpoint}}
\put(775.42,113.00){\usebox{\plotpoint}}
\put(796.18,113.00){\usebox{\plotpoint}}
\multiput(800,113)(20.756,0.000){0}{\usebox{\plotpoint}}
\put(816.93,113.00){\usebox{\plotpoint}}
\put(837.69,113.00){\usebox{\plotpoint}}
\multiput(841,113)(20.756,0.000){0}{\usebox{\plotpoint}}
\put(858.44,113.00){\usebox{\plotpoint}}
\put(879.20,113.00){\usebox{\plotpoint}}
\multiput(882,113)(20.756,0.000){0}{\usebox{\plotpoint}}
\put(896,113){\usebox{\plotpoint}}
\end{picture}
}
\bigskip \bigskip

%\hbox{\input{jadee2plot.tex} \input{durhamplot.tex}}
\hbox{
% GNUPLOT: LaTeX picture
\setlength{\unitlength}{0.240900pt}
\ifx\plotpoint\undefined\newsavebox{\plotpoint}\fi
\begin{picture}(974,720)(0,0)
\font\gnuplot=cmr10 at 10pt
\gnuplot
\sbox{\plotpoint}{\rule[-0.200pt]{0.400pt}{0.400pt}}%
\put(220.0,113.0){\rule[-0.200pt]{166.221pt}{0.400pt}}
\put(220.0,113.0){\rule[-0.200pt]{0.400pt}{129.845pt}}
\put(220.0,113.0){\rule[-0.200pt]{4.818pt}{0.400pt}}
\put(198,113){\makebox(0,0)[r]{0}}
\put(890.0,113.0){\rule[-0.200pt]{4.818pt}{0.400pt}}
\put(220.0,190.0){\rule[-0.200pt]{4.818pt}{0.400pt}}
\put(198,190){\makebox(0,0)[r]{50}}
\put(890.0,190.0){\rule[-0.200pt]{4.818pt}{0.400pt}}
\put(220.0,267.0){\rule[-0.200pt]{4.818pt}{0.400pt}}
\put(198,267){\makebox(0,0)[r]{100}}
\put(890.0,267.0){\rule[-0.200pt]{4.818pt}{0.400pt}}
\put(220.0,344.0){\rule[-0.200pt]{4.818pt}{0.400pt}}
\put(198,344){\makebox(0,0)[r]{150}}
\put(890.0,344.0){\rule[-0.200pt]{4.818pt}{0.400pt}}
\put(220.0,421.0){\rule[-0.200pt]{4.818pt}{0.400pt}}
\put(198,421){\makebox(0,0)[r]{200}}
\put(890.0,421.0){\rule[-0.200pt]{4.818pt}{0.400pt}}
\put(220.0,498.0){\rule[-0.200pt]{4.818pt}{0.400pt}}
\put(198,498){\makebox(0,0)[r]{250}}
\put(890.0,498.0){\rule[-0.200pt]{4.818pt}{0.400pt}}
\put(220.0,575.0){\rule[-0.200pt]{4.818pt}{0.400pt}}
\put(198,575){\makebox(0,0)[r]{300}}
\put(890.0,575.0){\rule[-0.200pt]{4.818pt}{0.400pt}}
\put(220.0,652.0){\rule[-0.200pt]{4.818pt}{0.400pt}}
\put(198,652){\makebox(0,0)[r]{350}}
\put(890.0,652.0){\rule[-0.200pt]{4.818pt}{0.400pt}}
\put(220.0,113.0){\rule[-0.200pt]{0.400pt}{4.818pt}}
\put(220,68){\makebox(0,0){0}}
\put(220.0,632.0){\rule[-0.200pt]{0.400pt}{4.818pt}}
\put(319.0,113.0){\rule[-0.200pt]{0.400pt}{4.818pt}}
\put(319,68){\makebox(0,0){0.05}}
\put(319.0,632.0){\rule[-0.200pt]{0.400pt}{4.818pt}}
\put(417.0,113.0){\rule[-0.200pt]{0.400pt}{4.818pt}}
\put(417,68){\makebox(0,0){0.1}}
\put(417.0,632.0){\rule[-0.200pt]{0.400pt}{4.818pt}}
\put(516.0,113.0){\rule[-0.200pt]{0.400pt}{4.818pt}}
\put(516,68){\makebox(0,0){0.15}}
\put(516.0,632.0){\rule[-0.200pt]{0.400pt}{4.818pt}}
\put(614.0,113.0){\rule[-0.200pt]{0.400pt}{4.818pt}}
\put(614,68){\makebox(0,0){0.2}}
\put(614.0,632.0){\rule[-0.200pt]{0.400pt}{4.818pt}}
\put(713.0,113.0){\rule[-0.200pt]{0.400pt}{4.818pt}}
\put(713,68){\makebox(0,0){0.25}}
\put(713.0,632.0){\rule[-0.200pt]{0.400pt}{4.818pt}}
\put(811.0,113.0){\rule[-0.200pt]{0.400pt}{4.818pt}}
\put(811,68){\makebox(0,0){0.3}}
\put(811.0,632.0){\rule[-0.200pt]{0.400pt}{4.818pt}}
\put(910.0,113.0){\rule[-0.200pt]{0.400pt}{4.818pt}}
\put(910,68){\makebox(0,0){0.35}}
\put(910.0,632.0){\rule[-0.200pt]{0.400pt}{4.818pt}}
\put(220.0,113.0){\rule[-0.200pt]{166.221pt}{0.400pt}}
\put(910.0,113.0){\rule[-0.200pt]{0.400pt}{129.845pt}}
\put(220.0,652.0){\rule[-0.200pt]{166.221pt}{0.400pt}}
\put(89,382){\makebox(0,0){\scriptsize B($y_3$)}}
\put(565,23){\makebox(0,0){ \scriptsize $y_3$ }}
\put(565,697){\makebox(0,0){$e^+e^- \to$ 3 jets\,\, Jade E-scheme}}
\put(220.0,113.0){\rule[-0.200pt]{0.400pt}{129.845pt}}
\put(220,113){\usebox{\plotpoint}}
\put(220.0,113.0){\rule[-0.200pt]{9.395pt}{0.400pt}}
\put(259.0,113.0){\rule[-0.200pt]{0.400pt}{81.665pt}}
\put(259.0,452.0){\rule[-0.200pt]{4.818pt}{0.400pt}}
\put(279.0,452.0){\rule[-0.200pt]{0.400pt}{13.731pt}}
\put(279.0,509.0){\rule[-0.200pt]{4.818pt}{0.400pt}}
\put(299.0,509.0){\rule[-0.200pt]{0.400pt}{4.577pt}}
\put(299.0,528.0){\rule[-0.200pt]{4.818pt}{0.400pt}}
\put(319.0,523.0){\rule[-0.200pt]{0.400pt}{1.204pt}}
\put(319.0,523.0){\rule[-0.200pt]{4.577pt}{0.400pt}}
\put(338.0,523.0){\rule[-0.200pt]{0.400pt}{4.336pt}}
\put(338.0,541.0){\rule[-0.200pt]{4.818pt}{0.400pt}}
\put(358.0,473.0){\rule[-0.200pt]{0.400pt}{16.381pt}}
\put(358.0,473.0){\rule[-0.200pt]{4.818pt}{0.400pt}}
\put(378.0,473.0){\rule[-0.200pt]{0.400pt}{0.482pt}}
\put(378.0,475.0){\rule[-0.200pt]{4.577pt}{0.400pt}}
\put(397.0,456.0){\rule[-0.200pt]{0.400pt}{4.577pt}}
\put(397.0,456.0){\rule[-0.200pt]{4.818pt}{0.400pt}}
\put(417.0,435.0){\rule[-0.200pt]{0.400pt}{5.059pt}}
\put(417.0,435.0){\rule[-0.200pt]{4.818pt}{0.400pt}}
\put(437.0,408.0){\rule[-0.200pt]{0.400pt}{6.504pt}}
\put(437.0,408.0){\rule[-0.200pt]{4.818pt}{0.400pt}}
\put(457.0,378.0){\rule[-0.200pt]{0.400pt}{7.227pt}}
\put(457.0,378.0){\rule[-0.200pt]{4.577pt}{0.400pt}}
\put(476.0,369.0){\rule[-0.200pt]{0.400pt}{2.168pt}}
\put(476.0,369.0){\rule[-0.200pt]{4.818pt}{0.400pt}}
\put(496.0,341.0){\rule[-0.200pt]{0.400pt}{6.745pt}}
\put(496.0,341.0){\rule[-0.200pt]{4.818pt}{0.400pt}}
\put(516.0,323.0){\rule[-0.200pt]{0.400pt}{4.336pt}}
\put(516.0,323.0){\rule[-0.200pt]{4.577pt}{0.400pt}}
\put(535.0,310.0){\rule[-0.200pt]{0.400pt}{3.132pt}}
\put(535.0,310.0){\rule[-0.200pt]{4.818pt}{0.400pt}}
\put(555.0,283.0){\rule[-0.200pt]{0.400pt}{6.504pt}}
\put(555.0,283.0){\rule[-0.200pt]{4.818pt}{0.400pt}}
\put(575.0,272.0){\rule[-0.200pt]{0.400pt}{2.650pt}}
\put(575.0,272.0){\rule[-0.200pt]{4.818pt}{0.400pt}}
\put(595.0,260.0){\rule[-0.200pt]{0.400pt}{2.891pt}}
\put(595.0,260.0){\rule[-0.200pt]{4.577pt}{0.400pt}}
\put(614.0,244.0){\rule[-0.200pt]{0.400pt}{3.854pt}}
\put(614.0,244.0){\rule[-0.200pt]{4.818pt}{0.400pt}}
\put(634.0,227.0){\rule[-0.200pt]{0.400pt}{4.095pt}}
\put(634.0,227.0){\rule[-0.200pt]{4.818pt}{0.400pt}}
\put(654.0,221.0){\rule[-0.200pt]{0.400pt}{1.445pt}}
\put(654.0,221.0){\rule[-0.200pt]{4.577pt}{0.400pt}}
\put(673.0,205.0){\rule[-0.200pt]{0.400pt}{3.854pt}}
\put(673.0,205.0){\rule[-0.200pt]{4.818pt}{0.400pt}}
\put(693.0,193.0){\rule[-0.200pt]{0.400pt}{2.891pt}}
\put(693.0,193.0){\rule[-0.200pt]{4.818pt}{0.400pt}}
\put(713.0,183.0){\rule[-0.200pt]{0.400pt}{2.409pt}}
\put(713.0,183.0){\rule[-0.200pt]{4.818pt}{0.400pt}}
\put(733.0,173.0){\rule[-0.200pt]{0.400pt}{2.409pt}}
\put(733.0,173.0){\rule[-0.200pt]{4.577pt}{0.400pt}}
\put(752.0,161.0){\rule[-0.200pt]{0.400pt}{2.891pt}}
\put(752.0,161.0){\rule[-0.200pt]{4.818pt}{0.400pt}}
\put(772.0,144.0){\rule[-0.200pt]{0.400pt}{4.095pt}}
\put(772.0,144.0){\rule[-0.200pt]{4.818pt}{0.400pt}}
\put(792.0,137.0){\rule[-0.200pt]{0.400pt}{1.686pt}}
\put(792.0,137.0){\rule[-0.200pt]{4.577pt}{0.400pt}}
\put(811.0,119.0){\rule[-0.200pt]{0.400pt}{4.336pt}}
\put(811.0,119.0){\rule[-0.200pt]{4.818pt}{0.400pt}}
\put(831.0,113.0){\rule[-0.200pt]{0.400pt}{1.445pt}}
\put(831.0,113.0){\rule[-0.200pt]{4.818pt}{0.400pt}}
\put(890,113){\usebox{\plotpoint}}
\put(890,113){\usebox{\plotpoint}}
\put(890.0,113.0){\rule[-0.200pt]{4.818pt}{0.400pt}}
\put(220,113){\usebox{\plotpoint}}
\put(220.00,113.00){\usebox{\plotpoint}}
\put(240.76,113.00){\usebox{\plotpoint}}
\multiput(259,113)(0.000,20.756){2}{\usebox{\plotpoint}}
\put(277.02,139.00){\usebox{\plotpoint}}
\multiput(279,139)(0.000,-20.756){0}{\usebox{\plotpoint}}
\put(288.78,130.00){\usebox{\plotpoint}}
\multiput(299,130)(0.000,-20.756){0}{\usebox{\plotpoint}}
\put(306.53,127.00){\usebox{\plotpoint}}
\multiput(319,127)(0.000,-20.756){0}{\usebox{\plotpoint}}
\put(326.29,126.00){\usebox{\plotpoint}}
\multiput(338,126)(0.000,-20.756){0}{\usebox{\plotpoint}}
\put(344.04,123.00){\usebox{\plotpoint}}
\put(364.80,123.00){\usebox{\plotpoint}}
\put(385.55,123.00){\usebox{\plotpoint}}
\multiput(397,123)(0.000,-20.756){0}{\usebox{\plotpoint}}
\put(404.31,121.00){\usebox{\plotpoint}}
\multiput(417,121)(0.000,20.756){0}{\usebox{\plotpoint}}
\put(423.07,123.00){\usebox{\plotpoint}}
\multiput(437,123)(0.000,-20.756){0}{\usebox{\plotpoint}}
\put(439.82,119.00){\usebox{\plotpoint}}
\multiput(457,119)(0.000,-20.756){0}{\usebox{\plotpoint}}
\put(459.58,118.00){\usebox{\plotpoint}}
\put(480.33,118.00){\usebox{\plotpoint}}
\multiput(496,118)(0.000,-20.756){0}{\usebox{\plotpoint}}
\put(500.09,117.00){\usebox{\plotpoint}}
\multiput(516,117)(0.000,-20.756){0}{\usebox{\plotpoint}}
\put(519.84,116.00){\usebox{\plotpoint}}
\put(540.60,116.00){\usebox{\plotpoint}}
\put(561.35,116.00){\usebox{\plotpoint}}
\put(582.11,116.00){\usebox{\plotpoint}}
\put(602.87,116.00){\usebox{\plotpoint}}
\put(623.62,116.00){\usebox{\plotpoint}}
\multiput(634,116)(0.000,-20.756){0}{\usebox{\plotpoint}}
\put(643.38,115.00){\usebox{\plotpoint}}
\put(664.13,115.00){\usebox{\plotpoint}}
\put(684.89,115.00){\usebox{\plotpoint}}
\put(705.64,115.00){\usebox{\plotpoint}}
\put(726.40,115.00){\usebox{\plotpoint}}
\put(747.15,115.00){\usebox{\plotpoint}}
\put(767.91,115.00){\usebox{\plotpoint}}
\multiput(772,115)(0.000,-20.756){0}{\usebox{\plotpoint}}
\put(787.66,114.00){\usebox{\plotpoint}}
\multiput(792,114)(0.000,20.756){0}{\usebox{\plotpoint}}
\put(806.42,116.00){\usebox{\plotpoint}}
\put(827.18,116.00){\usebox{\plotpoint}}
\multiput(831,116)(0.000,-20.756){0}{\usebox{\plotpoint}}
\put(845.93,114.00){\usebox{\plotpoint}}
\put(866.69,114.00){\usebox{\plotpoint}}
\multiput(871,114)(0.000,-20.756){0}{\usebox{\plotpoint}}
\put(886.44,113.00){\usebox{\plotpoint}}
\put(907.20,113.00){\usebox{\plotpoint}}
\put(910,113){\usebox{\plotpoint}}
\end{picture}
% GNUPLOT: LaTeX picture
\setlength{\unitlength}{0.240900pt}
\ifx\plotpoint\undefined\newsavebox{\plotpoint}\fi
\begin{picture}(974,720)(0,0)
\font\gnuplot=cmr10 at 10pt
\gnuplot
\sbox{\plotpoint}{\rule[-0.200pt]{0.400pt}{0.400pt}}%
\put(220.0,113.0){\rule[-0.200pt]{166.221pt}{0.400pt}}
\put(220.0,113.0){\rule[-0.200pt]{0.400pt}{129.845pt}}
\put(220.0,113.0){\rule[-0.200pt]{4.818pt}{0.400pt}}
\put(198,113){\makebox(0,0)[r]{0}}
\put(890.0,113.0){\rule[-0.200pt]{4.818pt}{0.400pt}}
\put(220.0,221.0){\rule[-0.200pt]{4.818pt}{0.400pt}}
\put(198,221){\makebox(0,0)[r]{20}}
\put(890.0,221.0){\rule[-0.200pt]{4.818pt}{0.400pt}}
\put(220.0,329.0){\rule[-0.200pt]{4.818pt}{0.400pt}}
\put(198,329){\makebox(0,0)[r]{40}}
\put(890.0,329.0){\rule[-0.200pt]{4.818pt}{0.400pt}}
\put(220.0,436.0){\rule[-0.200pt]{4.818pt}{0.400pt}}
\put(198,436){\makebox(0,0)[r]{60}}
\put(890.0,436.0){\rule[-0.200pt]{4.818pt}{0.400pt}}
\put(220.0,544.0){\rule[-0.200pt]{4.818pt}{0.400pt}}
\put(198,544){\makebox(0,0)[r]{80}}
\put(890.0,544.0){\rule[-0.200pt]{4.818pt}{0.400pt}}
\put(220.0,652.0){\rule[-0.200pt]{4.818pt}{0.400pt}}
\put(198,652){\makebox(0,0)[r]{100}}
\put(890.0,652.0){\rule[-0.200pt]{4.818pt}{0.400pt}}
\put(220.0,113.0){\rule[-0.200pt]{0.400pt}{4.818pt}}
\put(220,68){\makebox(0,0){0}}
\put(220.0,632.0){\rule[-0.200pt]{0.400pt}{4.818pt}}
\put(319.0,113.0){\rule[-0.200pt]{0.400pt}{4.818pt}}
\put(319,68){\makebox(0,0){0.05}}
\put(319.0,632.0){\rule[-0.200pt]{0.400pt}{4.818pt}}
\put(417.0,113.0){\rule[-0.200pt]{0.400pt}{4.818pt}}
\put(417,68){\makebox(0,0){0.1}}
\put(417.0,632.0){\rule[-0.200pt]{0.400pt}{4.818pt}}
\put(516.0,113.0){\rule[-0.200pt]{0.400pt}{4.818pt}}
\put(516,68){\makebox(0,0){0.15}}
\put(516.0,632.0){\rule[-0.200pt]{0.400pt}{4.818pt}}
\put(614.0,113.0){\rule[-0.200pt]{0.400pt}{4.818pt}}
\put(614,68){\makebox(0,0){0.2}}
\put(614.0,632.0){\rule[-0.200pt]{0.400pt}{4.818pt}}
\put(713.0,113.0){\rule[-0.200pt]{0.400pt}{4.818pt}}
\put(713,68){\makebox(0,0){0.25}}
\put(713.0,632.0){\rule[-0.200pt]{0.400pt}{4.818pt}}
\put(811.0,113.0){\rule[-0.200pt]{0.400pt}{4.818pt}}
\put(811,68){\makebox(0,0){0.3}}
\put(811.0,632.0){\rule[-0.200pt]{0.400pt}{4.818pt}}
\put(910.0,113.0){\rule[-0.200pt]{0.400pt}{4.818pt}}
\put(910,68){\makebox(0,0){0.35}}
\put(910.0,632.0){\rule[-0.200pt]{0.400pt}{4.818pt}}
\put(220.0,113.0){\rule[-0.200pt]{166.221pt}{0.400pt}}
\put(910.0,113.0){\rule[-0.200pt]{0.400pt}{129.845pt}}
\put(220.0,652.0){\rule[-0.200pt]{166.221pt}{0.400pt}}
\put(89,382){\makebox(0,0){\scriptsize B($y_D$)}}
\put(565,23){\makebox(0,0){ \scriptsize $y_D$ }}
\put(565,697){\makebox(0,0){$e^+e^- \to$ 3 jets\,\, $k_\perp$-algorithm}}
\put(220.0,113.0){\rule[-0.200pt]{0.400pt}{129.845pt}}
\put(220,113){\usebox{\plotpoint}}
\put(220.0,113.0){\rule[-0.200pt]{4.818pt}{0.400pt}}
\put(240.0,113.0){\rule[-0.200pt]{0.400pt}{62.875pt}}
\put(240.0,374.0){\rule[-0.200pt]{4.577pt}{0.400pt}}
\put(259.0,374.0){\rule[-0.200pt]{0.400pt}{41.194pt}}
\put(259.0,545.0){\rule[-0.200pt]{4.818pt}{0.400pt}}
\put(279.0,545.0){\rule[-0.200pt]{0.400pt}{7.227pt}}
\put(279.0,575.0){\rule[-0.200pt]{4.818pt}{0.400pt}}
\put(299.0,563.0){\rule[-0.200pt]{0.400pt}{2.891pt}}
\put(299.0,563.0){\rule[-0.200pt]{4.818pt}{0.400pt}}
\put(319.0,553.0){\rule[-0.200pt]{0.400pt}{2.409pt}}
\put(319.0,553.0){\rule[-0.200pt]{4.577pt}{0.400pt}}
\put(338.0,517.0){\rule[-0.200pt]{0.400pt}{8.672pt}}
\put(338.0,517.0){\rule[-0.200pt]{4.818pt}{0.400pt}}
\put(358.0,494.0){\rule[-0.200pt]{0.400pt}{5.541pt}}
\put(358.0,494.0){\rule[-0.200pt]{4.818pt}{0.400pt}}
\put(378.0,485.0){\rule[-0.200pt]{0.400pt}{2.168pt}}
\put(378.0,485.0){\rule[-0.200pt]{4.577pt}{0.400pt}}
\put(397.0,425.0){\rule[-0.200pt]{0.400pt}{14.454pt}}
\put(397.0,425.0){\rule[-0.200pt]{4.818pt}{0.400pt}}
\put(417.0,425.0){\rule[-0.200pt]{0.400pt}{5.782pt}}
\put(417.0,449.0){\rule[-0.200pt]{4.818pt}{0.400pt}}
\put(437.0,385.0){\rule[-0.200pt]{0.400pt}{15.418pt}}
\put(437.0,385.0){\rule[-0.200pt]{4.818pt}{0.400pt}}
\put(457.0,377.0){\rule[-0.200pt]{0.400pt}{1.927pt}}
\put(457.0,377.0){\rule[-0.200pt]{4.577pt}{0.400pt}}
\put(476.0,359.0){\rule[-0.200pt]{0.400pt}{4.336pt}}
\put(476.0,359.0){\rule[-0.200pt]{4.818pt}{0.400pt}}
\put(496.0,355.0){\rule[-0.200pt]{0.400pt}{0.964pt}}
\put(496.0,355.0){\rule[-0.200pt]{4.818pt}{0.400pt}}
\put(516.0,324.0){\rule[-0.200pt]{0.400pt}{7.468pt}}
\put(516.0,324.0){\rule[-0.200pt]{4.577pt}{0.400pt}}
\put(535.0,289.0){\rule[-0.200pt]{0.400pt}{8.431pt}}
\put(535.0,289.0){\rule[-0.200pt]{4.818pt}{0.400pt}}
\put(555.0,289.0){\rule[-0.200pt]{0.400pt}{2.650pt}}
\put(555.0,300.0){\rule[-0.200pt]{4.818pt}{0.400pt}}
\put(575.0,271.0){\rule[-0.200pt]{0.400pt}{6.986pt}}
\put(575.0,271.0){\rule[-0.200pt]{4.818pt}{0.400pt}}
\put(595.0,270.0){\usebox{\plotpoint}}
\put(595.0,270.0){\rule[-0.200pt]{4.577pt}{0.400pt}}
\put(614.0,240.0){\rule[-0.200pt]{0.400pt}{7.227pt}}
\put(614.0,240.0){\rule[-0.200pt]{4.818pt}{0.400pt}}
\put(634.0,237.0){\rule[-0.200pt]{0.400pt}{0.723pt}}
\put(634.0,237.0){\rule[-0.200pt]{4.818pt}{0.400pt}}
\put(654.0,220.0){\rule[-0.200pt]{0.400pt}{4.095pt}}
\put(654.0,220.0){\rule[-0.200pt]{4.577pt}{0.400pt}}
\put(673.0,208.0){\rule[-0.200pt]{0.400pt}{2.891pt}}
\put(673.0,208.0){\rule[-0.200pt]{4.818pt}{0.400pt}}
\put(693.0,189.0){\rule[-0.200pt]{0.400pt}{4.577pt}}
\put(693.0,189.0){\rule[-0.200pt]{4.818pt}{0.400pt}}
\put(713.0,186.0){\rule[-0.200pt]{0.400pt}{0.723pt}}
\put(713.0,186.0){\rule[-0.200pt]{4.818pt}{0.400pt}}
\put(733.0,175.0){\rule[-0.200pt]{0.400pt}{2.650pt}}
\put(733.0,175.0){\rule[-0.200pt]{4.577pt}{0.400pt}}
\put(752.0,152.0){\rule[-0.200pt]{0.400pt}{5.541pt}}
\put(752.0,152.0){\rule[-0.200pt]{4.818pt}{0.400pt}}
\put(772.0,139.0){\rule[-0.200pt]{0.400pt}{3.132pt}}
\put(772.0,139.0){\rule[-0.200pt]{4.818pt}{0.400pt}}
\put(792.0,121.0){\rule[-0.200pt]{0.400pt}{4.336pt}}
\put(792.0,121.0){\rule[-0.200pt]{4.577pt}{0.400pt}}
\put(811.0,113.0){\rule[-0.200pt]{0.400pt}{1.927pt}}
\put(890,113){\usebox{\plotpoint}}
\put(890,113){\usebox{\plotpoint}}
\put(890.0,113.0){\rule[-0.200pt]{4.818pt}{0.400pt}}
\put(220,113){\usebox{\plotpoint}}
\put(220.00,113.00){\usebox{\plotpoint}}
\put(240.00,113.76){\usebox{\plotpoint}}
\put(251.51,123.00){\usebox{\plotpoint}}
\put(272.27,123.00){\usebox{\plotpoint}}
\multiput(279,123)(0.000,20.756){0}{\usebox{\plotpoint}}
\put(291.02,125.00){\usebox{\plotpoint}}
\multiput(299,125)(0.000,-20.756){0}{\usebox{\plotpoint}}
\put(308.78,122.00){\usebox{\plotpoint}}
\put(329.53,122.00){\usebox{\plotpoint}}
\multiput(338,122)(0.000,-20.756){0}{\usebox{\plotpoint}}
\put(348.29,120.00){\usebox{\plotpoint}}
\multiput(358,120)(0.000,20.756){0}{\usebox{\plotpoint}}
\put(368.04,121.00){\usebox{\plotpoint}}
\put(388.80,121.00){\usebox{\plotpoint}}
\multiput(397,121)(0.000,20.756){0}{\usebox{\plotpoint}}
\put(406.55,124.00){\usebox{\plotpoint}}
\put(427.31,124.00){\usebox{\plotpoint}}
\multiput(437,124)(0.000,-20.756){0}{\usebox{\plotpoint}}
\put(445.07,121.00){\usebox{\plotpoint}}
\put(465.82,121.00){\usebox{\plotpoint}}
\multiput(476,121)(0.000,20.756){0}{\usebox{\plotpoint}}
\put(485.58,122.00){\usebox{\plotpoint}}
\multiput(496,122)(0.000,-20.756){0}{\usebox{\plotpoint}}
\put(505.33,121.00){\usebox{\plotpoint}}
\multiput(516,121)(0.000,20.756){0}{\usebox{\plotpoint}}
\put(525.09,122.00){\usebox{\plotpoint}}
\put(545.84,122.00){\usebox{\plotpoint}}
\multiput(555,122)(0.000,-20.756){0}{\usebox{\plotpoint}}
\put(564.60,120.00){\usebox{\plotpoint}}
\put(585.35,120.00){\usebox{\plotpoint}}
\put(606.11,120.00){\usebox{\plotpoint}}
\put(626.87,120.00){\usebox{\plotpoint}}
\put(647.62,120.00){\usebox{\plotpoint}}
\multiput(654,120)(0.000,20.756){0}{\usebox{\plotpoint}}
\put(667.38,121.00){\usebox{\plotpoint}}
\multiput(673,121)(0.000,-20.756){0}{\usebox{\plotpoint}}
\put(687.13,120.00){\usebox{\plotpoint}}
\put(707.89,120.00){\usebox{\plotpoint}}
\multiput(713,120)(0.000,20.756){0}{\usebox{\plotpoint}}
\put(727.64,121.00){\usebox{\plotpoint}}
\multiput(733,121)(0.000,-20.756){0}{\usebox{\plotpoint}}
\put(746.40,119.00){\usebox{\plotpoint}}
\multiput(752,119)(0.000,-20.756){0}{\usebox{\plotpoint}}
\put(766.15,118.00){\usebox{\plotpoint}}
\put(786.91,118.00){\usebox{\plotpoint}}
\multiput(792,118)(0.000,20.756){0}{\usebox{\plotpoint}}
\put(806.66,119.00){\usebox{\plotpoint}}
\put(827.42,119.00){\usebox{\plotpoint}}
\multiput(831,119)(0.000,-20.756){0}{\usebox{\plotpoint}}
\put(846.18,117.00){\usebox{\plotpoint}}
\multiput(851,117)(0.000,-20.756){0}{\usebox{\plotpoint}}
\put(864.93,115.00){\usebox{\plotpoint}}
\multiput(871,115)(0.000,-20.756){0}{\usebox{\plotpoint}}
\put(884.69,114.00){\usebox{\plotpoint}}
\multiput(890,114)(0.000,-20.756){0}{\usebox{\plotpoint}}
\put(904.44,113.00){\usebox{\plotpoint}}
\put(910,113){\usebox{\plotpoint}}
\end{picture}
}

\caption{Coefficient of $(\alpha_s/(2\pi))^2$ for the thrust,
C-parameter distributions and the $y_{\rm cut}$ distributions of the
Jade E and $k_\perp$-clustering algorithms. The thrust distribution is
multiplied by ($1-t$), the C-parameter distribution is multiplied by
$C$, while the distributions of the clustering algorithms are 
multiplied by $y_{\rm cut}$. The dotted histograms show the
statistical errors.}

\end{figure}

\section{Conclusion}

In this paper we have presented a simple generalization of the
subtraction method of ref.\ \cite{KSjets} for the calculation of any
infrared safe physical cross section in perturbative QCD. The apparent 
conflict between the need for important sampling in order to achieve
sufficient numerical precision and the increasing difficulty of
performing partial fractioning in the tree-level \NLO matrix element
was overcome by a decomposition of the phase space such that in one
region only one Lorentz invariant of the external momenta can become
singular in the calculation of an $N$-jet observable. We wish to
emphasize the simplicity of our algorithm: the necessary analytic
integrals are rather trivial, while the numerical implementation is 
only a little more complicated than a tree-level Monte Carlo program.

We have given all the necessary integrals that define any \NLO QCD jet
cross section explicitly (see eqs.\ (\ref{IfinA}), (\ref{Ifinf}) and 
(\ref{INbody})). The phase space integrations in these integrals can be
programmed for any number of jets. Once having such a master program the
only ingredients that have to be changed in a modular fashion are the
\begin{itemize}
\item the Born-level and \NLO tree matrix elements in four dimensions
($\Psi^{(2n)}$ ($n=N,N+1$) functions);
\item the color linked Born matrix elements ($\Psi^{(2N;c)}$ functions);
\item the non-singular part of the one-loop helicity amplitudes
($\Am_{{\rm NS}}$ functions);
\item the $\S_n$ ($n=N,N+1$) measurement functions.
\end{itemize}

The algorithm can be trivially altered for calculating QCD jet cross
sections in other processes, like in $e^+e^-$ annihilation or
deep-inelastic scattering. As an example, we have shown results of the
\NLO thrust, C-parameter and jet distributions in the case of $e^+e^-$
annihilation.  The numerical convergence was found to be similar to that
of the program of ref. \cite{KN3jet} in the case of $e^+e^-$ annihilation
into three jets. In the case of three-jet production in hadron collisions
such a benchmark calculation does not exist yet. In order to demonstrate
the applicability of the algorithm in such calculations, in a companion paper
\cite{pp3jetg}, we give results of a \NLO calculation of three-jet cross
section in hadron collisions in the simplified case of pure Yang-Mills
theory. The structure of the algorithm is essentially the same when quarks
are included therefore, the conclusions are expected to be similar
in the full QCD case.

\bigskip
\large {\bf Acknowledgment:}\normalsize \quad 
One of us (Z.T.) is grateful to Z. Kunszt for a number of useful discussions.
This research was supported in part by the EEC Programme "Human Capital
and Mobility", Network "Physics at High Energy Colliders", contract
PECO ERBCIPDCT 94 0613 as well as by the Foundation for Hungarian Higher 
education and Research, the Hungarian National Science Foundation grant
OTKA T-016613 and the Research Group in Physics of the Hungarian Academy
of Sciences, Debrecen.

\newpage

\Large 
\noindent {\bf Appendix}
\normalsize
\renewcommand{\theequation}{A.\arabic{equation}}
\setcounter{equation}{0}

\bigskip

This appendix is added for the readers' convenience. It contains a 
collection of the integrals that were used in the main text. The explicit
evaluation of these integrals is rather trivial, therefore, minimal 
details are given.

First we calculate the soft integral
\beqn
&&\J_{mn}(\vec{p}) = \int 
(2\pi\mu)^{2\ve}\d^{4-2\ve} p_{N+1}\,2\,\delta\!\left(p_{N+1}^2\right) 
\\ \nn &&\qquad\qquad\quad \times 
\frac{2\s mn}{\s m{,N+1}(\s m{,N+1}+\s n{,N+1})}
\Theta(\alpha\s mn > \s m{,N+1}+\s n{,N+1}).
\eeqn
for $m,n=1,\ldots,N$, $m\ne n$. The integrand depends only on Lorentz 
invariants therefore, the integral can be calculated in any frame. 
We choose the ``$m$-$n$'' system, where the four-momenta
of gluon $m$, $n$ and $N+1$ take the form (first four components 
are energy, $z$, $x$ and $y$ components of the three-momentum)
\beqn
&&p_m^\mu = \frac{\sqrt{\s mn}}{2}(1,1,0,0,\ldots), \\
\vspaceinarray
&&p_n^\mu = \frac{\sqrt{\s mn}}{2}(1,-1,0,0,\ldots), \\
\vspaceinarray
&&p_{N+1}^\mu = E(1,\cos \vartheta,\sin \vartheta \cos \varphi,
\sin \vartheta \sin \varphi,\ldots),
\eeqn
where the dots mean $d-4$ zeros. With this choice the invariants are
\beqn
&&\s m{,N+1} = \sqrt{\s mn} E (1-\cos \vartheta), \\
\vspaceinarray
&&\s n{,N+1} = \sqrt{\s mn} E (1+\cos \vartheta). 
\eeqn
Consequently, in the ``$m$-$n$'' system the integral takes the following
simple form:
\beq
\J_{mn} = \int \left(\frac{2\pi\mu}{E}\right)^{2\ve}
\frac{\d E}{E}\,(\sin \vartheta)^{-2\ve}
\d\vartheta \,(\sin \varphi)^{-2\ve}\d \varphi \,\d^{-2\ve}\Omega
\frac{\sin \vartheta}{1-\cos\vartheta}\Theta(\alpha\sqrt{\s mn}/2>E).
\eeq
This integral is easily evaluated and one obtains the exact result:
\beq
\J_{mn} = \frac{\pi}{\ve^2}\left(\frac{4\pi\mu^2}{\s mn}\right)^\ve
\alpha^{-2\ve}\frac{\Gamma(1-\ve)}{\Gamma(1-2\ve)}.
\eeq

In the main text, we have also used the following integrals: 
\beq
\label{Q2phi}
\int \left(\frac{Q^2}{E_P^2}\right)^{-\ve} \frac{\d Q^2}{Q^2}
\Theta(4Q_{\rm max}^2>Q^2)
\,(\sin \varphi)^{-2\ve}\d \varphi \,\d^{-2\ve}\Omega
=-\frac{2\pi}{\ve}Q_{\rm max}^{-2\ve}\frac{(4\pi)^{-\ve}}{\Gamma(1-\ve)},
\eeq
\beqn
\label{Zam}
&&\Z_a(\alpha) = \int^1 \d z\,(1-z)^{-2\ve}
\left[z^{-2\ve} \sum_b \P_{b/a}(z)\Theta(z>1/2)
-\frac{2C(a)}{1-z}\Theta(z>1-\alpha)\right]\qquad
\\ \nn &&\qquad\quad
= -\gamma(a)-2C(a)\ln \alpha 
+\ve\gamma'(a)+2\ve C(a)\left(\ln^2 \alpha+\frac{\pi^2}{3}\right),
\eeqn
where
\beq
\gamma'(g) = -\frac{67}{9}\Nc + \frac{23}{18}\Nf,
\qquad \gamma'(q) = -\frac{13}{4}\frac{V}{\Nc}
\eeq
and
\beq
\label{Wint}
(2\pi\mu)^{2\ve} \int \frac{\d^{2-2\ve}{\bf W}}{W^2} \Theta(W_{\rm max}>W)
=-\frac{1}{\ve}\frac{\pi}{\Gamma(1-\ve)}
\left(\frac{4\pi\mu^2}{W_{\rm max}^2}\right)^\ve,
\eeq
The derivation of these results is sufficiently simple so that we can
omit the details.

\def\np#1#2#3  {Nucl.\ Phys.\ #1 (19#3) #2}
\def\pl#1#2#3  {Phys.\ Lett.\ #1 (19#3) #2}
\def\prep#1#2#3  {Phys.\ Rep.\ #1 (19#3) #2}
\def\pr#1#2#3 {Phys.\ Rev.\ #1 (19#3) #2}
\def\prl#1#2#3 {Phys.\ Rev.\ Lett. #1 (19#3) #2}
\def\zp#1#2#3  {Zeit.\ Phys.\ Lett. #1 (19#3) #2}
\def\cmc#1#2#3  {Comp.\ Phys.\ Comm. #1 (19#3) #2}


\baselineskip=18pt

\begin{thebibliography}{99}
\bibitem{EKSjets} S. D. Ellis, Z. Kunszt, D. Soper, \pr{D40}{2188}{89} .
\bibitem{KSjets} Z. Kunszt, D. Soper, \pr{D46}{192}{92} .
\bibitem{GGepem}W. T. Giele and E. W. N Glover, \pr{D46}{1980}{92} .
\bibitem{GGKpp}W. T. Giele, E. W. N Glover and D. A. Kosower,
\np{B403}{633}{93} .
\bibitem{factorization} A. Bassetto, M. Ciafaloni and G. Marchesini,
\prep{100}{202}{83} ;\\
G. Altarelli and G. Parisi, \np{B126}{298}{77} .
\bibitem{CSdipol} S. Catani and M. H. Seymour, preprint CERN-TH/96-28,
hep-ph/9602277.
\bibitem{EKSincl} S. D. Ellis, Z. Kunszt, D. Soper, \prl{62}{726}{88} ;
S. D. Ellis, Z. Kunszt, D. Soper, \prl{64}{2121}{90} .
\bibitem{EKS2jet} S. D. Ellis, Z. Kunszt, D. Soper, \prl{69}{1496}{92} .
\bibitem{MP90} M. Mangano and T. Parke, \prep{200}{301}{90} .
\bibitem{BDK5g} Z. Bern L. Dixon, and D. A. Kosower, \prl{70}{2677}{93} .
\bibitem{KST4q1g} Z.Kunszt, A. Signer and Z. Tr\'ocs\'anyi,
\pl{B336}{529}{94} .
\bibitem{BDK2q3g} Z. Bern L. Dixon, and D. A. Kosower, \np{B437}{259}{95} .
\bibitem{FKSjets} S. Frixione, Z. Kunszt, A. Signer, preprint
ETH-TH/95-42. %hep-ph/95122XX.
\bibitem{KSTsing} Z. Kunszt, A. Signer and Z. Tr\'ocs\'anyi, 
\np{B420}{550}{94} .
\bibitem{KST2to2} 
Z. Kunszt, A. Signer and Z. Tr\'ocs\'anyi, \np{B411}{397}{94} .
\bibitem{mueller} For a review, see J. C. Collins, G. Sterman and D. E.
Soper, ``Factorization of hard processes in QCD'' in A. H. Mueller, ed.,
Perturbative Quantum Chromodynamics, (World Scientific, Singapore, 1989).
\bibitem{ES} R. K. Ellis and J. Sexton, \np{B269}{445}{86} .
\bibitem{BKcolor} Z. Bern and D. A. Kosower, \np{B362}{389}{91} .
\bibitem{BDKcoll} Z. Bern L. Dixon and D. A. Kosower, \np{B425}{217}{94} .
\bibitem{rambo} S. D. Ellis, R. Kleiss and W. J. Stirling, \cmc{40}{359}{86} .
\bibitem{KN3jet}  Z. Kunszt and P. Nason,  ``QCD'' form
Z Physics at LEP1, CERN 89-08 p.\ 434
\bibitem{ERT} R. K. Ellis, D. A. Ross and A. E. Terrano,
\np{B178}{421}{81} .
\bibitem{pp3jetg} Z. Tr\'ocs\'anyi, Three-jet cross section in hadron
collisions at \NLO: pure gluon processes, Preprint KLTE-DTP/96-2.
\end{thebibliography}
\end{document}